\documentclass{bmcart}

\usepackage[utf8]{inputenc} %unicode support
\usepackage[left]{lineno}
\usepackage{amssymb,latexsym}
\usepackage[mathscr]{eucal}
\usepackage{setspace}
\usepackage{bbold}
\usepackage{subcaption}
\usepackage{graphicx}
\usepackage{amsmath}
\usepackage{amsfonts}
\usepackage{amsbsy}
\usepackage[mathscr]{eucal}
\usepackage{amsthm}
\usepackage{esint}
\usepackage{stix}

\def\mE{\mathbb{E}}

\def\citep{\cite}

\def\bs{\boldsymbol}
\DeclareMathOperator*{\E}{\mathbb{E}}
\def\bm{\boldsymbol}

\startlocaldefs
\endlocaldefs

\begin{document}

%%% Start of article front matter
\begin{frontmatter}

\begin{fmbox}
%\dochead{Methodology Article}


\title{\bf A State Space Technique for Wildlife Position Estimation Using Non-Simultaneous Signal Strength Measurements}

%%%%%%%%%%%%%%%%%%%%%%%%%%%%%%%%%%%%%%%%%%%%%%
%%                                          %%
%% Enter the authors here                   %%
%%                                          %%
%% Specify information, if available,       %%
%% in the form:                             %%
%%   <key>={<id1>,<id2>}                    %%
%%   <key>=                                 %%
%% Comment or delete the keys which are     %%
%% not used. Repeat \author command as much %%
%% as required.                             %%
%%                                          %%
%%%%%%%%%%%%%%%%%%%%%%%%%%%%%%%%%%%%%%%%%%%%%%

\author[
   addressref={aff1},                   % id's of addresses, e.g. {aff1,aff2}
        corref={aff1},   
   % id of corresponding address, if any
   %noteref={aff1},                        % id's of article notes, if any
   email={janaswamy@ecs.umass.edu, $^2$pamela.h.loring@eco.umass.edu, $^3$JamesMcLaren@cunet.Carleton.ca}   % email address
]{\inits{RJ}\fnm{Ramakrishna} \snm{Janaswamy}}
 \author[
   addressref={aff2},
   email={pamela.h.loring@eco.umass.edu}
]{\inits{PL}\fnm{Pamela} \snm{Loring}}
\author[
   addressref={aff3},
%email={jmclaren@udel.edu}
email={JamesMcLaren@cunet.Carleton.ca}
]{\inits{JM}\fnm{James D.} \snm{McLaren}}

%%%%%%%%%%%%%%%%%%%%%%%%%%%%%%%%%%%%%%%%%%%%%%
%%                                          %%
%% Enter the authors' addresses here        %%
%%                                          %%
%% Repeat \address commands as much as      %%
%% required.                                %%
%%                                          %%
%%%%%%%%%%%%%%%%%%%%%%%%%%%%%%%%%%%%%%%%%%%%%%

\address[id=aff1]{%                           % unique id
  \orgname{University of Massachusetts, Department of Electrical \& Computer Engineering}, % university, etc
  \street{100 Natural Resources Road},                     %
 % \postcode{01003}                                % post or zip code
  \city{Amherst, MA 01003},                              % city
  \cny{USA}
  }                                   % country  
\address[id=aff2]{%
  \orgname{US Fish and Wildlife Service, Division of Migratory Birds Northeast Region},
  \street{50 Bend Road},
 % \postcode{01003}
  \city{Charlestown, RI 02813},
  \cny{USA}
}
\address[id=aff3]{%
  \orgname{National Wildlife Research Center},
  \street{Environment and Climate Change Canada},
  \city{Ottawa}
  \postcode{K1A0H3},
  \cny{Canada;}
   \orgname{Guest Researcher, Biology Department},
  \street{Carleton University},
  \city{Ottawa}
  \postcode{K1S5B6},
  \cny{Canada}
}
%%%%%%%%%%%%%%%%%%%%%%%%%%%%%%%%%%%%%%%%%%%%%%
%%                                          %%
%% Enter short notes here                   %%
%%                                          %%
%% Short notes will be after addresses      %%
%% on first page.                           %%
%%                                          %%
%%%%%%%%%%%%%%%%%%%%%%%%%%%%%%%%%%%%%%%%%%%%%%

\begin{artnotes}
%\note{Sample of title note}     % note to the article
%\note[id=n1]{Equal contributor} % note, connected to author
\end{artnotes}

\end{fmbox}% comment this for two column layout

%%%%%%%%%%%%%%%%%%%%%%%%%%%%%%%%%%%%%%%%%%%%%%
%%                                          %%
%% The Abstract begins here                 %%
%%                                          %%
%% Please refer to the Instructions for     %%
%% authors on http://www.biomedcentral.com  %%
%% and include the section headings         %%
%% accordingly for your article type.       %%
%%                                          %%
%%%%%%%%%%%%%%%%%%%%%%%%%%%%%%%%%%%%%%%%%%%%%%

\begin{abstractbox}

\begin{abstract} % abstract
\parttitle{Background} %if any
Fixed very-high-frequency (VHF) antenna arrays have been a standard technique for remote tracking of radio-tagged wildlife since the 1960s. In recent years, a growing network of coordinated, fixed VHF arrays on a shared frequency has expanded throughout the Western Hemisphere, but our ability to estimate animal trajectories at regional scales is limited by the fact that detections consist of irregular time series of signal strength values as tagged individuals move in and out of detection range of antenna beams within the array. Therefore more advanced modeling methods are needed to refine estimates of animal locations. A novel state-space technique is presented to estimate the location and airborne movements of VHF tagged wildlife individuals with fixed VHF arrays. The approach combines a movement model (Ornstein- Uhlenbeck random process in the transverse (horizontal) plane and a Cox- Ingersoll-Ross process in the vertical direction) to ensure biologically-consistent trajectories in three-dimensions, and an observation model to account for the effect of range, altitude and bearing angle on the received signal strength. The observation model of received signals accounts for low-end saturation from receiver noise, high-end saturation from receiver non-linearities as well as a wireless multipath phenomena, which modulates the received signal according to range, the altitude and radiation characteristics of the Yagi array. A pattern function for the Yagi array is synthesized that facilitates linearization of the received signals and subsequent application of Kalman filtering.  

\parttitle{Results} %if any
We first validate the model using a simulated trajectory and then estimate the space-time trajectory of a migrating VHF-tagged shorebird, which was tracked with a regional automated radio telemetry network. The algorithm accurately predicted the average movement trajectory given the system parameters and the initial conditions (average error $<$ 1 km). The modeled shorebird track represents a first estimate in three-dimensional (3D) of a radio tagged bird using a fixed telemetry array, and was qualitatively reasonable, but exhibited some sensitivity in the vertical plane and to initial conditions.

\parttitle{Conclusions}
From an applied ecology perspective, we feel that our technique is flexible and broadly applicable to other telemetry networks and ecological systems. Thus, given appropriate array configuration and modification of the observation and process models, our technique could be applied to the analysis of automated VHF telemetry data across a wide range of spatial and temporal scales, from site-specific studies using targeted arrays, to coordinated digital VHF tracking efforts that span the Hemisphere.
\end{abstract}

%%%%%%%%%%%%%%%%%%%%%%%%%%%%%%%%%%%%%%%%%%%%%%
%%                                          %%
%% The keywords begin here                  %%
%%                                          %%
%% Put each keyword in separate \kwd{}.     %%
%%                                          %%
%%%%%%%%%%%%%%%%%%%%%%%%%%%%%%%%%%%%%%%%%%%%%%


% MSC classifications codes, if any
%\begin{keyword}[class=AMS]
%\kwd[Primary ]{}
%\kwd{}
%\kwd[; secondary ]{}
%\end{keyword}

\end{abstractbox}
%
%\end{fmbox}% uncomment this for twcolumn layout

\end{frontmatter}

%%%%%%%%%%%%%%%%%%%%%%%%%%%%%%%%%%%%%%%%%%%%%%
%%                                          %%
%% The Main Body begins here                %%
%%                                          %%
%% Please refer to the instructions for     %%
%% authors on:                              %%
%% http://www.biomedcentral.com/info/authors%%
%% and include the section headings         %%
%% accordingly for your article type.       %%
%%                                          %%
%% See the Results and Discussion section   %%
%% for details on how to create sub-sections%%
%%                                          %%
%% use \cite{...} to cite references        %%
%%  \cite{koon} and                         %%
%%  \cite{oreg,khar,zvai,xjon,schn,pond}    %%
%%  \nocite{smith,marg,hunn,advi,koha,mouse}%%
%%                                          %%
%%%%%%%%%%%%%%%%%%%%%%%%%%%%%%%%%%%%%%%%%%%%%%

%%%%%%%%%%%%%%%%%%%%%%%%% start of article main body
% <put your article body there>

%%%%%%%%%%%%%%%%
%% Background %%
%%
\def\Di{\Delta_i}
\def\sigY{\sigma_{_Y}}
\section{Background}

Accurate estimation and assessment of animal movements is an essential tool towards understanding and conserving animal populations and ecosystem processes \cite{Newton2008, Ponchon2013, hays2016}, \citep{Nathan2008}. Radio telemetry has been a standard technique for tracking wildlife since the 1960s \citep{Cochran1965}, and to date remains one of the sole options for collecting individual-based tracking data of small-bodied ($<$ 100\,g) species \citep{Ponchon2013}. For example, accurate positioning of birds fitted with light-weight ($\ge 0.25$\,g) VHF transmitters is possible using handheld receivers; manual tracking is however limited by effort, available transportation and knowledge of an animal's whereabouts following release \citep{Cochran1965, Thorup2007, schmal2011}. 

The recent expansion of coordinated digital VHF telemetry using automated radio telemetry systems has enabled a wide range of wildlife taxa to be tracked at unprecedented spatial and temporal scales. Automated radio telemetry systems can provide round-the-clock detection for relatively long periods (e.g. 4+ months for 1-g units with a 5-second pulse rate) whenever tagged individuals are within detection range of receiving antennas \citep{Kenward1987, Larkin1996}. Such systems have enabled studies to be conducted in a wide range of environments, from grassland \citep{Tucker2014} to tropical rainforest \citep{Kays2011}, to track a variety of taxa including reptiles \citep{Ward2013, Sperry2013}, birds \citep{Mitchell2012, Leyrer2006, Smolinsky2013, Smetzer2017, gomez2017}, terrestrial mammals \citep{Enzo2008}, and bats \citep{McGuire2012}. These studies typically base location on triangulation via implied bearings, estimated in turn via relative signal strengths between multi-antenna array elements. More recently, more accurate methods have been developed using time of arrival and high-precision clocks 
\citep{Weiser}, \citep{maccurdy}. These studies have all been applied to ranges within 5-20 km and animals either on the ground or presumed to be flying at low altitudes. Contrastingly, a coordinated network of automated radio telemetry stations, the Motus Wildlife Tracking System, was piloted in northeastern North America in 2012 and has since expanded throughout western Hemisphere (www.motus-wts.org). These stations have recorded over 250 million detections from more than 9,000 individuals representing 87 species of birds, bats, and insects tracked with digital VHF transmitters on a shared frequency \citep{Taylor2017}. However, given the data collected by automated radio telemetry stations typically consists of unevenly-spaced time series of signal strength values received by single beams or sporadically by multiple antenna beams, estimating the true positions and trajectories of radio-tagged animals involving changes in flight altitude at regional scales remains a quantitatively complex and an unresolved area of research \citep{cochran1972, Taylor2017}. 
\section{Methods}
\subsection{Aim, Design and Setting}
\label{sec:art}
The objective of our study is to provide a framework to estimate the position of radio-tagged birds based on time-series data logged by an automated radio telemetry array, which accounts for the 3D structure. In this study, we used digitally coded VHF transmitters (NTQB-4-2; Lotek Wireless; hereafter nanotag) that weighed 1\,g and measured $12 \times 8\times 8$\, mm, with an external 18\,cm long wire antenna.  Each transmitter was programmed to emit signals at 166.38\,MHz on a pulse duration of 5.3\,s, from activation through the end of battery life (approximately 170 days). 
In collaboration with the Motus Wildlife Tracking System, we established an array of automated of sixteen radio telemetry towers positioned across a greater than150\,km section of the U.S. Atlantic coast from Cape Cod, MA to Long Island, NY (see Fig.~\ref{fig:PamFig1}). Each tower consisted of six Yagi antennas (Cushcraft PLC, 9-element Yagi antenna), each having a horizontal plane beam-width of 35$^o$ and whose main beams were separated in bearing by 60$^o$ and end-mounted in a radial configuration atop a 12.2\,m mast (see Fig.~\ref{fig:PamFig2}).  The antennas were connected via a coaxial cable to ports on a data logging receiving unit (SRX600, Lotek Wireless) that ran continuously using two 12-V deep-cycle batteries charged by a 120-Watt solar system.

Each receiving unit was programmed to automatically log several types of data from each antenna, including:  transmitter ID number, date, time (synchronized among all receivers in network using Global Positioning System (GPS) clocks), antenna (defined by receiving station and bearing), and power received (linear display scale:  0-255).  The received time-series power signal was sampled sequentially by each Yagi antenna of a tower, each with a dwell time of 6.5\,s (greater than the pulse duration). Thus the minimum separation between two successive readings at any Yagi antena is 32.5\,s. However, the time-series detection data can be non-uniformly spaced as tagged individuals traverse regions of the antenna beam from which signal strength is very low and not detected by the receivers. \par

We collected systematic calibration data of the reception pattern of the automated telemetry stations using a test transmitter attached to a partially frozen bird carcass that was suspended from a kite and flown from the back of a boat \citep{Taylor2011}.  During our calibration surveys, the boat was traveled at a constant speed of approximately 10\,m/s and the transmitter flew at an altitude of approximately 30\,m. 

To estimate the bird location based on these measurements, we develop a state-space based Kalman filtering technique, consisting of a system movement model and an observation model. The state-space modeling approach is similar to that employed by \citep{Johnson2008, albertsen2015}, but the movement model was extended to accommodate correlated vertical motion based on a Cox-Ingersoll-Ross model \citep{cir1985} in addition to having correlated motion in the transverse plane, modeled by a modified Ornstein-Uhlenbeck velocity model \citep{uhlenbeck1930}. Conditions required for the dynamical system to be observable (uniquely solvable) are discussed. The observation model is further tailored to a Yagi antenna pattern conforming to the manufacturer's catalog data. We conducted calibrations of signal strength measurements of the receiver to develop a linearized observation model that facilitates estimation by standard Kalman filtering.  We then test the validity of the Kalman filter model on a simulated trajectory, and lastly apply the algorithm to predict the trajectory of a VHF-tagged shorebird tracked over several hours by an automated radio telemetry array along the U.S Atlantic coast.\par

Our paper is organized as follows. The radio telemetry system is described in \S\ref{sec:art} and challenges specific to measurement system are described in \S\ref{sec:ChO}. The movement model in the transverse plane and vertical direction are first developed in \S\ref{sec:transmod}. The Yagi antenna pattern used in the observation model is constructed in \S\ref{sec:Yagi}, and making use of calibrated kite measurements, the receiver component governing signal gain is developed in \S\ref{sec:rxmod}. The linearized observation model for estimation by standard Kalman filtering is then developed in \S\ref{sec:linmod} and the Kalman filter algorithm is presented in \S\ref{sec:Kal}. The validity of the models so developed is then tested in \S\ref{sec:simul} on a controlled, synthetic bird trajectory. Finally, in \S\ref{sec:estimated}, the algorithm is applied to predict the migratory bird trajectory using measurements gathered by our array of receivers. 

\subsection{Challenges and Observations}
\label{sec:ChO}
\begin{figure}[h!]
\centering
  \begin{subfigure}[htb]{0.5\textwidth}
    \includegraphics[width=\textwidth]{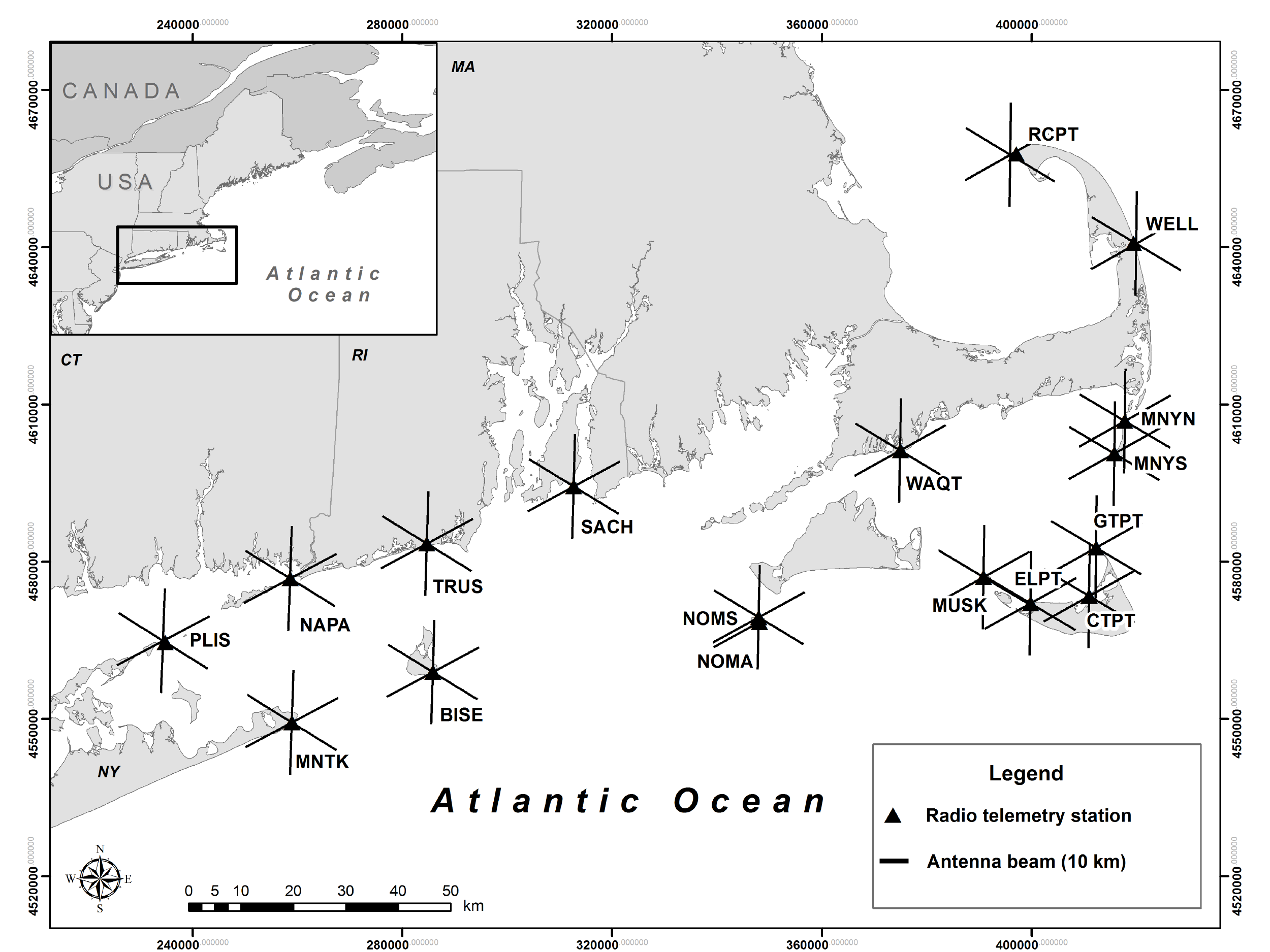}
    \caption{Study area map showing geometry of 16 tracking towers distributed along the U. S. Atlantic Coast from Cape Cod, MA to Long Island, NY.}
    \label{fig:PamFig1}
  \end{subfigure}
  \begin{subfigure}[htb]{0.65\textwidth}
    \includegraphics[width=\textwidth]{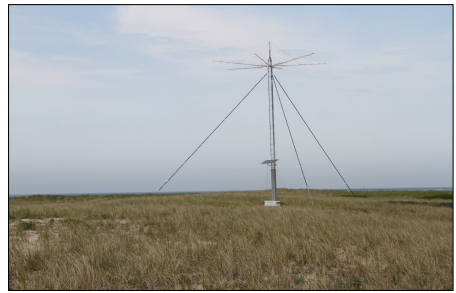}
    \caption{A 12.2 m tower supporting six PLC-1669 Yagi arrays.}
    \label{fig:PamFig2}
  \end{subfigure}
  \caption{Deployment of towers and photograph of a tower antenna array. Each radial\\ line in the left figure represents the axis of a Yagi array. The array axes are separated by\\ 60$^o$ and the horizontal plane beamwidth is each array is $35^o$.} 
  \label{fg:Tower}
\end{figure}

The sixteen towers that were erected in the Atlantic ocean region for the purpose of this study are shown in Fig.~\ref{fig:PamFig1}. Fig.~\ref{fig:PamFig2} shows the typical set of 6 Yagi arrays erected at each tower. Given the relatively small size of the nano-tag at the operating wavelength, it is reasonable to first assume that the radio nano-tags have an omnidirectional pattern. 

Our measurement system and received data are characterized by the following challenges and observations which influence our model development: 
\begin{enumerate}
\item{Ch$_1$:} Simultaneous measurements over multiple towers were not available, thus precluding triangulation. 
\item{Ch$_2$:} The receiver records integer values in the range $Z\in(0,255)$ based on the power received on a decibel scale from proprietary coded signals. 
\item{Ch$_3$:} The power received at any given time is not explicitly dependent on the instantaneous speed of nano-tag. Thus bird speed cannot be directly inferred from power measurements alone. 
\item{Ob$_1$:} When a signal is recorded by the receiver attached to a tower, it merely implies that the bird is within its range. Here, we are trying, in addition, to estimate the location of the bird in 3D space based on the intensity of the received signal. Assuming identical tags, the power received depends in a complicated way on the combined location coordinates of the moving transmitter and the radiation pattern of the Yagi array. The same signal power can be received from a multitude of locations even if the transmitter remains in the main beam of the receiving antenna. There is a good probability of even receiving the signals via the antenna side lobes and back lobe. Thus the inversion of three spatial coordinates from a single power measurement is necessarily non-unique. 
\item{Ob$_2$:} Signal was received over highly non-uniform time-intervals. The minimum separation between successive measurements at any tower was around 5\,s. The maximum separation between successive observations could involve multiple towers and exceeded over 12 hours in our case. 
\item{Ob$_3$:} It was observed experimentally that during certain times the bird was detected via multiple beams and towers that were widely separated.   This would be possible if the bird were flying at high altitudes. So a dynamic height dependence must be included in the theoretical models. 
\end{enumerate}

It might be tempting to consider a simple dynamic model consisting of three {\em independent local level} \citep{DK2012} systems along the three spatial coordinates. However, a quick analysis will reveal that it will render the system unobservable, thereby, resulting in highly ill-conditioned inversions. {\em Observability\/} means that the state vector can be uniquely determined from finite number of measurements \citep{rugh1996}. One way to address the non-observability quandary is to consider a first-order system with decay and select varying decay constants in the three-spatial directions. In addition to suffering from anisotropic behavior in space, such a first order system will result in very limited translational motion in space. We therefore consider a first-order system of the Ornstein-Uhlenbeck type in which it is the speed variable (and not the position variable) that driven by white noise. The corresponding displacement will be equal to the time integral of speed and will consist of a correlated random walk with a good component of drift.  Our movement and observation models are characterized by the following specific features:
\begin{enumerate}
\item 
For the sake of analysis, the bird is treated like a particle that is described by its three spatial coordinates and any applicable speeds. The movement models in the transverse $(x,y)$ plane  and in the vertical coordinate, $z$, are different, somewhat mimicking the true flight pattern of a shorebird. In particular, the model permits non-rectilinear motion in the transverse plane and level motion with fluctuations in the vertical plane such that altitude always remains above the mean sea level.  Additionally, model parameters governing flight can be optimized to reflect either directed or highly irregular movement.
\item In the transverse plane, the state space along each coordinate consists of location and speed as state variables. In the vertical coordinate the state space consists of location variable only. 
\item The bird movement model has the required complexity to make all of state variables {\em almost observable\/} despite the receiver power being dependent only on the location coordinates as highlighted in observation $Ob_1$. Making the system observable will also address the challenge $Ch_3$ by coupling the speed variables implicitly to power measurements.  
\item The vertical coordinate comes into play both through the range (to tower) coordinate as well as through the multipath phenomenon. Presence of multipath will  account for the height gain trend that is highlighted in observation $Ob_3$. To make the dynamic model more physically reasonable we make the transverse coordinates and the vertical coordinate statistically correlated. 
\item The overall system has the parameters tuned such that the model is more-or-less isotropic in the transverse plane. The resulting equations in the final discrete time model are not subject to any approximations even for irregularly spaced measurements in time. 
\item A receiver model is constructed that directly relates the received power to the display reading $Z$. This is based on {\em non-linear} least square fitting of  calibration data. This will address the challenge highlighted under $Ch_2$. Saturation at the upper end caused by receiver non-linearity and at the lower end by receiver noise is accounted for in the receiver model by the incorporation of a soft-limiter and measurement noise respectively. \label{recmod}
\item An analytical antenna pattern is synthesized that mimics the radiation pattern supplied by the manufacturer.
\item A non-linear observation model is constructed from an analytical antenna pattern function, range, height-gain factor and the receiver model as determined in item (\ref{recmod}) above. 
\item Non-linear Kalman filtering technique \cite{DSimon2006} is utilized to determine the {\em mean} trajectory of the bird based on the system model assumed and conditioned by the measurements. 
\item When large time gaps in the received data are encountered, the model is initialized again with the initial coordinates for the new run set such that they are closest to the most recently determined mean coordinates. The overall mean trajectory is sensitive to the initial coordinates assigned to the model as well as to the strategy employed for handling large time gaps. In coming up with schemes for dealing with large time gaps, we keep in mind the absolute maximum speed that the bird can fly to narrow down on the many possibilities. 
\end{enumerate}
\subsection{Correlated Motion Model}
\label{sec:transmod}
\begin{figure}[h!]
\centerline{\scalebox{0.375}{\includegraphics{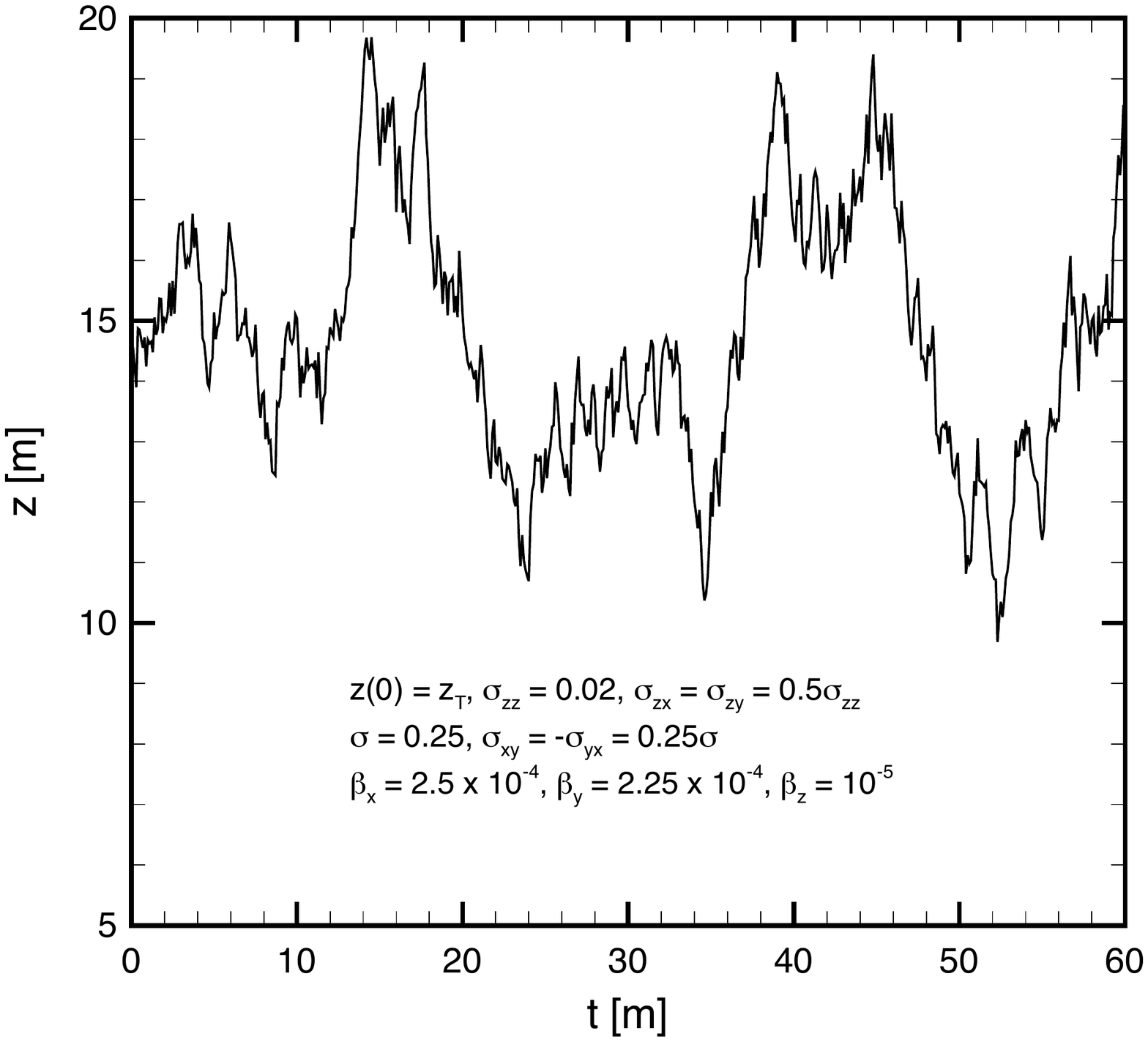}}}
\caption{Sample instantaneous vertical coordinate over a 60\,minute time period in the\\ presence of system noise. $z_T = 14.72\,$m.}
\label{fg:VertFig}
\end{figure}

The dynamical variables in the transverse plane are the particle position, $(x(t), y(t))$, and the corresponding instantaneous velocity components $(v_x(t), v_y(t))$, whereas in the vertical plane the dynamical variable is $x_z(t)$ such that the altitude coordinate is $z(t) = x_z^2(t)>0$. We denote time derivative by an over-dot on the variable. The equations of motion are governed by an Ornstein-Uhlenbeck (OU) process \citep{uhlenbeck1930}, \citep{mahnke2009} without drift in the transverse plane and by the  Cox-Ingersoll-Ross model \citep{cir1985} in the vertical direction: 
\begin{eqnarray}
\dot{x}(t) &=&v_x(t)\label{eq:Apsys1}\\[0.5ex]
\dot{v}_x(t)&=&-\beta_xv_x(t)+\sigma_{xx}W_x(t)+\sigma_{xy}W_y(t)\label{eq:Apsys2}\\[0.5ex]
\dot{y}(t) &=&v_y(t)\label{eq:Apsys3}\\[0.5ex]
\dot{v}_y(t)&=&-\beta_yv_y(t)+\sigma_{yx}W_x(t)+\sigma_{yy}W_y(t)\label{eq:Apsys4}\\[0.5ex]
\dot{x}_z(t) &=& -\beta_z x_z(t)  +\sigma_{zx}W_x(t) + \sigma_{zy}W_y(t)+\sigma_{zz} W_z(t),\label{eq:vert1}
\end{eqnarray}
where $(\beta_x, \beta_y, \beta_z)$ are positive real constants with units of [$s^{-1}$], which determine the decay of $(v_x(t), v_y(t), x_z(t))$, $W_x(t)$, $W_y(t)$ and $W_z(t)$ are white Gaussian noise processes\citep{papoulis2002} with the property that $\E[W_x(t)] = 0, \E[W_x(t_1)W_x(t_2)]= \delta(t_2-t_1)$, $\delta(\cdot)$ being the Dirac delta function and likewise for $W_y(t)$, and $W_z(t)$, $\sigma_{xx}, \sigma_{yy}, \sigma_{zz}$ are positive constants with units [ms$^{-3/2}$] that determine the variances of white noise and $\sigma_{xy}, \sigma_{yx}, \sigma_{zx}, \sigma_{zy}$ are real constants that determine the correlation between $v_x$, $v_y$ and $x_z$. It is assumed that the noise processes $W_x(t)$ and $W_y(t)$ driving the two Cartesian components are independent of each other and are also independent of $W_z(t)$. The operation $\E(\cdot)$ denotes ensemble average and will involve all stochastic quantities. The first order equation (\ref{eq:Apsys2}) satisfied by the velocity variable $v_x(t)$ with a decay constant $\beta_x$ and driven by random forces $\sigma_{xx}W_x(t)$ and $\sigma_{xy}W_y(t)$ is simply a statement of Newton's force law. 
It is possible to incorporate environmental covariates such as wind speed and direction \citep{Johnson2008} into the present movement model by way of introducing potentials and the corresponding external forces\footnote{For a particle of mass $m$ with the potential energy, $U_P(x;t)$, the force per unit mass arising from the potential is -$\frac{1}{m}\frac{\partial U_P}{\partial x}$.} into the right hand side of (\ref{eq:Apsys2}). However, we will not consider external forces here. The first equation (\ref{eq:Apsys1}) relates the time rate of change of position to velocity. A {\em fixed point\/} of the system in the absence of noise is obtained by setting all time derivatives to zero. In the $x$-coordinate this results in $(x(t),v_x(t)) = (c, 0)$ for the model above, where $c$ is a constant. Fixed points are points of stagnation in the sense that if the state trajectory reaches it at some instant of time, it will remain there for all future times in the absence of noise. In an ordinary OU process  a constant term $\beta_x\gamma_0$ is also added to the right hand side of the velocity equation (\ref{eq:Apsys2}). In that case the velocity settles around $v_x = \gamma_0$ and the position will asymptotically reach $\gamma_0t$ in the absence of noise. Consequently, the transverse trajectories will tend be rectilinear in an ordinary OU process in the absence of noise. To remove this unwanted bias and to model non-directed movement, we assume $\gamma_0 = 0$ here. For simplicity of notation, we will omit the argument $t$ from all quantities when there is no danger of confusion. Using It\^o's formula \citep{karat1998}, \citep{oksen2003} it can be shown that the altitude satisfies the non-linear dynamical equation
\begin{eqnarray}
\dot{z}(t) &=& -2\beta_zz(t) +2\sqrt{z(t)}[\sigma_{zx}W_x(t) + \sigma_{zy}W_y(t) + \sigma_{zz}W_z(t)]\nonumber\\
&& +\ (\sigma_{zx}^2+\sigma^2_{zy}+\sigma^2_{zz}).\label{eq:vert3}
\end{eqnarray}
In mathematical finance, a model such as this is referred to as the Cox-Ingersoll-Ross model \citep{cir1985} that describes the evolution of interest rates (assuming they are $>0$) over time. The fixed point for altitude is $z(\infty) = (\sigma_{zx}^2+\sigma_{zy}^2 + \sigma_{zz}^2)/2\beta_z=:z_\infty$. 
Figure~\ref{fg:VertFig} shows a sample evolution of the altitude $z(t)$ for $\sigma_{zz} = 0.02\,$m$^{1/2}$s$^{-1/2}$, $z(0) = 14.72$\,m and $z(\infty) = 30\,$m. 
It is seen that the model permits fine-scale variation in altitude to capture the altitude variation of birds in flight. Because of the non-linearity present in (\ref{eq:vert3}), fluctuations in the height variable $z(t)$ will be more correlated than those of the intermediate variable $x_z(t)$. To maintain isotropy of coupling between $z$ to $x$ and $z$ to $y$ we choose $\sigma_{zx}=\sigma_{zy}$.  
Denoting the state vector $\bs{p} = [x,v_x, y, v_y,x_z]'$, where a prime denotes matrix transpose and the noise vector by ${\bs w} = [W_x,W_y, W_z]'$, the dynamical equations can be written in a matrix form as 
\begin{equation}
\dot{\bs p} = {\bf T}{\bs p} + {\bf B}{\bs w},\label{eq:state}\end{equation}
where  the block-diagonal system matrix $\bf T$ is 
 \begin{equation}
 {\bf T} =
 \begin{bmatrix}
 0&1&&&\\
 0&-\beta_x&&&\\
 &&0&1&&\\
 &&0&-\beta_y&\\
 &&&&-\beta_z
 \end{bmatrix}\end{equation}
 
and the noise coefficient matrix $\bf B$ is 

\begin{equation}
 {\bf B} = \begin{bmatrix}
 0&0&0\\
 \sigma_{xx}&\sigma_{xy}&0\\
 0&0&0\\
 \sigma_{yx}&\sigma_{yy}&0\\
 \sigma_{zx}&\sigma_{zy}&\sigma_{zz}
 \end{bmatrix}.\label{eq:sysmat} \end{equation}
The  parameters contained in the system are $[\beta_x,\sigma_{xx},\beta_y,\sigma_{yy},\beta_z,\sigma_{zz}, \sigma_{xy}, \sigma_{yx}, \sigma_{zx}, \sigma_{zy}]$ and are ten in number.  Equation (\ref{eq:state}) represents a linear time-invariant (LTI) system driven by random white Gaussian noise. Even though only the velocity components of the state vector and $x_z$ are directly driven by noise, the noise gets coupled to the other component by virtue of the system matrix ${\bf T}$ being non-diagonal. However, because of linearity and time-invariance of the system, it can be shown that the noise present in the components remains white Gaussian, albeit correlated. System (\ref{eq:state}) can be solved by employing techniques of stochastic calculus \citep{oksen2003} or by using the traditional concept of transition matrix and Laplace transforms \citep{luen1979}, \citep{rugh1996}. Using any of these approaches, a recursive solution to (\ref{eq:state}) can be obtained as
\begin{equation}
\bs{p}(t_i+\Di) = e^{{\bf T}\Delta_i}\bs{p}(t_i) + \bs{\nu}_i\label{eq:sol}\end{equation}
where
\begin{equation}
e^{{\bf T}t}=\left[\begin{matrix}
1& \mu_{x}(t)&&&\\[0.5ex]
0&\lambda_{x}(t)&&&\\[0.5ex]
&&1&\mu_{y}(t)&\\[0.5ex]
&&0&\lambda_{y}(t)&\\[0.5ex]
&&&&\lambda_{z}(t)
\end{matrix}
\right],\end{equation}

where $\lambda_{k}(t) = \exp(-\beta_kt)$, $\mu_k(t) = (1-\lambda_{k})/\beta_k$, $k=x,y,z$ and ${\bs\nu}_i$ is a zero mean, white Gaussian random vector with a covariance matrix $\bs{Q}_i$:
\begin{equation}
\bs{Q}_i=\E({\bs \nu}_i{\bs\nu}_j') = \delta_i^j\int\limits_0^{\Di} e^{{\bf T}(\Di-\tau_i)}{\bf BI}_3{\bf B}'e^{{\bf T}'(\Di-\tau_i)}\,d\tau_i\label{eq:cov1}
\end{equation}
where $\delta_i^j$ is the Kronecker's delta and ${\bf I}_3$ is an identity matrix of size $3$. Note that $\mu_k(t)\to t$ and $\lambda_k(t) = 1$ as $\beta_k \to 0,\ k = x,y,z$. It is also clear that $\bs{Q}_i$ is a Hermitian matrix, {\em viz.,} $\bs{Q}_i(n,m) = \left[\bs{Q}_i(m,n)\right]^*$, where superscript * denotes complex conjugation. 
The entries of the matrix $Q_i$ can be easily found as the integrals involved are elementary. For instance,
\begin{equation}
Q_i(1,1) = \frac{a[\Di - 2\mu_x(\Di)]}{\beta_x^2}+\frac{Q_i(2,2)]}{\beta_x^2};\quad Q_i(2,2) = \frac{a[1-\lambda_x^2(\Di)]}{2\beta_x},\end{equation}
where $a = \sigma_{xx}^2+\sigma^2_{xy}$. Equation (\ref{eq:sol}) determines the state vector at time $t_i+\Di$ given the state vector and the stochastic input $\bs{\nu}_i$ at time $t_i$. It is an exact equation in the sense that it is not subject to any discretization error and does not require the time increment $\Di$ to be uniform. To gain some insight into the parameters $\sigma_{xx}, \sigma_{xy}$ and $\beta_{x}$, let us consider the recursive solution of the velocity equation (\ref{eq:Apsys2}):
\begin{equation}
v_x(t_i+\Di) = v_x(t_i)e^{-\beta_{x}\Di} + \sqrt{Q_i(2,2)}\, {\cal N}(0,1),\label{eq:vx}\end{equation}
where ${\cal N}(0,1)$ is a normal random variable. 

For $\beta_{x}\Di\ll 1$, $Q_i(2,2) \sim a\Di$ and  
\begin{equation}
v_x(t_i+\Di) \approx v_x(t_i) + \sqrt{Q_i(2,2)}\, {\cal N}(0,1).\label{eq:vxs}\end{equation}
In this case, the updated velocity is relatively independent of the parameter $\beta_x$ and jumps from its previous value by a Gaussian random variable of standard deviation $\sqrt{a\Di}$. This regime will lead to substantial displacements (integral of velocity) and may be a good choice for describing migration type of bahavior. 

On the other hand for $\beta_{x}\Di\gg 1, v_x(t_i) = O(1)$ and ${a/\beta_x}\ge 1$, we have 
\begin{equation}
v_x(t_i+\Di)\approx \sqrt{\frac{a}{2\beta_{x}}}\,{\cal N}(0,1).\label{eq:vxl}
\end{equation}
In this case the updated velocity is independent of the previous velocity and is mostly determined by random fluctuations and the parameter $\beta_x$. As a result the displacement here will be small and this choice may be most suitable for describing nesting type of behavior. Here we may think of $\Di = 1/\beta_x$ as 
the correlation time for the velocity component $v_x$. For instance, when $\beta_x = 2.5\times 10^{-4}$ [s$^{-1}$] the correlation time for velocity is about 1.1 hour. \par
Similar conclusions may be drawn for $\beta_z, \sigma_{zz}, \sigma_{zx}=\sigma_{zy}$ by noting from (\ref{eq:vert1}) that 
\begin{equation}
x_z(t_i+\Di) = x_z(t_i)\,e^{-\beta_z\Di} + \sqrt{Q_i(5,5)}\, {\cal N}(0,1),\quad Q_i(5,5) = \left[1-\lambda_z^2(\Di)\right]z_\infty,\label{eq:xz}
\end{equation}
where $z_\infty$ is defined after (\ref{eq:vert3}). 

\begin{figure}[h!]
\centering
  \begin{subfigure}[htb]{0.475\textwidth}
    \includegraphics[width=\textwidth]{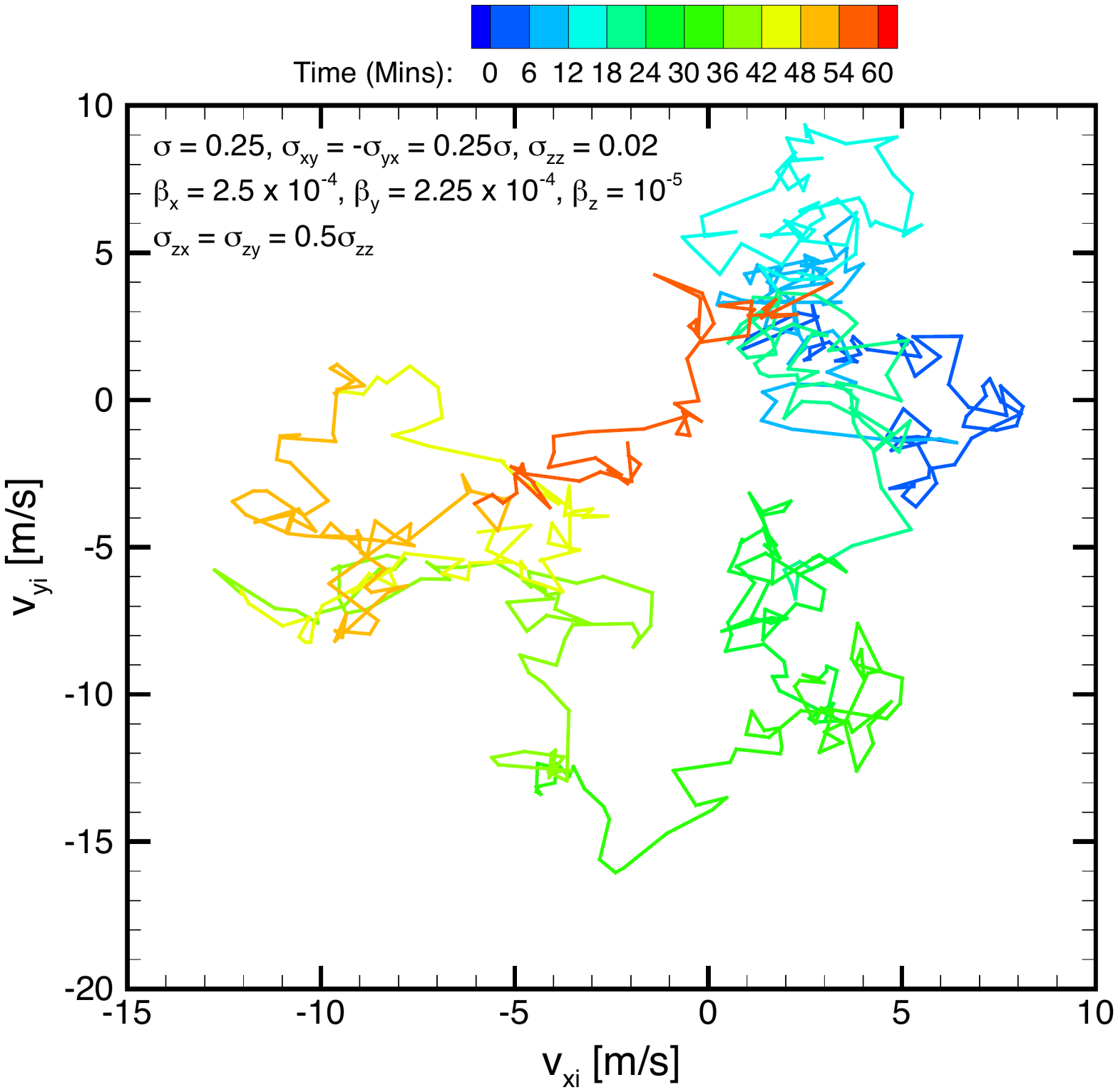}
    \caption{Instantaneous velocity}
    \label{fig:trans1}
  \end{subfigure}
  \begin{subfigure}[htb]{0.475\textwidth}
    \includegraphics[width=\textwidth]{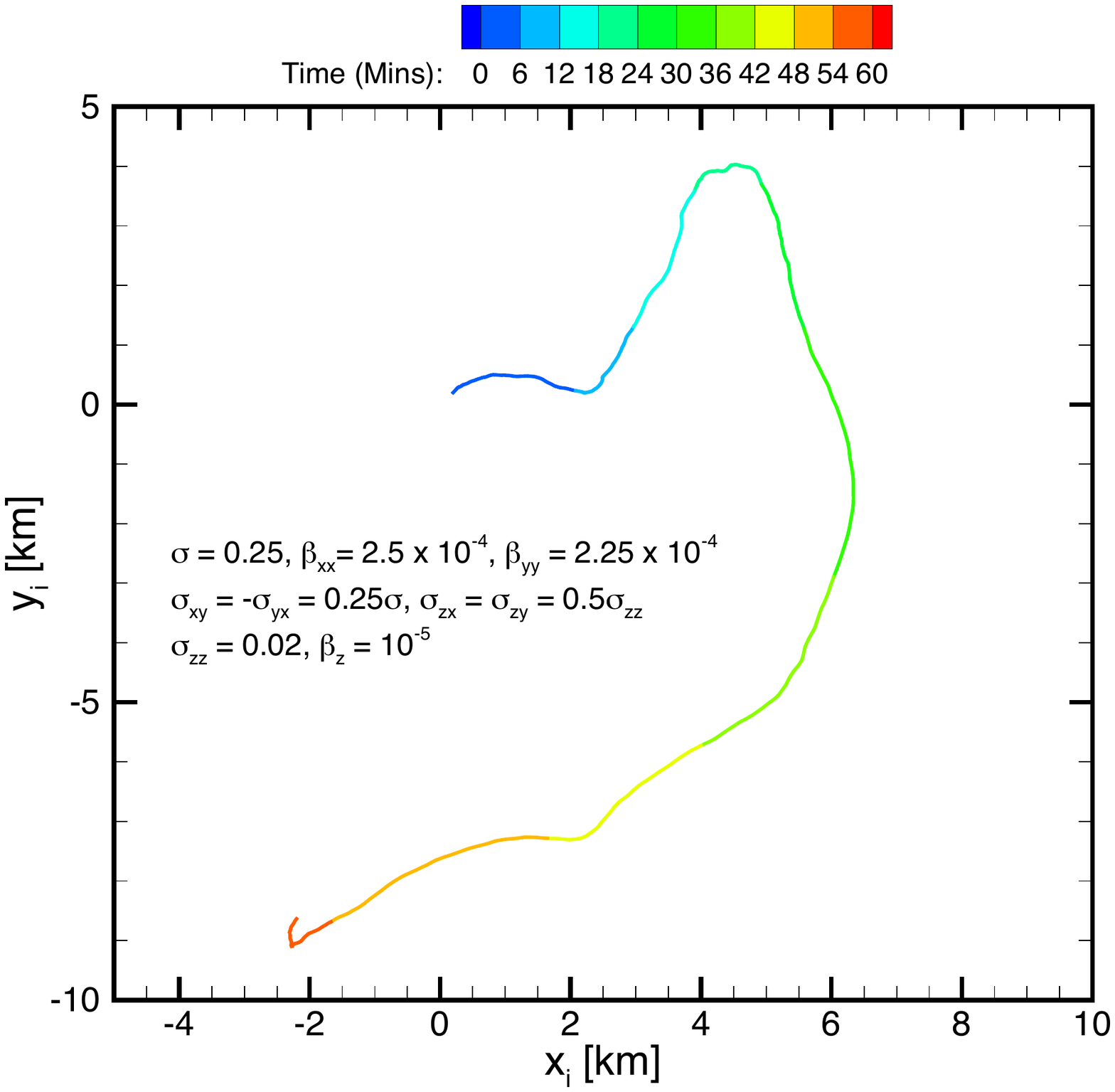}
    \caption{Instantaneous position}
    \label{fig:trans2}
  \end{subfigure}
  \caption{Sample instantaneous velocity and instantaneous position in the transverse\\ plane in the presence of system noise.}
  \label{fg:transtraj}
\end{figure}

It is desirable to keep the pairs $(\beta_x,\sigma_{xx})$ and $(\beta_y,\sigma_{yy})$ close to each other and choose $\sigma_{xy}, \sigma_{yx}$ appropriately so as to maintain {\em isotropy} of motion. To understand this, let the polar coordinates corresponding to $ (v_x(t),v_y(t))$ be $(v(t),\theta(t))$. Using $v_x(t) = v(t)\cos\theta(t), v_y(t) = v(t)\sin\theta(t)$, it is easy to see that
\begin{eqnarray}
\left[\begin{matrix}\dot{v}\\ v\dot{\theta}\end{matrix}\right]  &=& \left[\begin{matrix}\cos\theta(t)&\sin\theta(t)\\-\sin\theta(t)&\cos\theta(t)\end{matrix}\right]\left\{\left[
\begin{matrix}
-\beta_x&0\\ 0&-\beta_y\end{matrix}\right]\left[\begin{matrix}v_x\\ v_y\end{matrix}\right]\right. \nonumber\\
&&\left.+ \left[\begin{matrix}\sigma_{xx}&\sigma_{xy}\\ \sigma_{yx}&\sigma_{yy}\end{matrix}\right]\left[\begin{matrix}W_x(t)\\ W_y(t)\end{matrix}\right]\right\}\nonumber\\ [1ex]
&=:& \left[\begin{matrix}\cos\theta(t)&\sin\theta(t)\\-\sin\theta(t)&\cos\theta(t)\end{matrix}\right]\left[
\begin{matrix}
-\beta_x&0\\ 0&-\beta_y\end{matrix}\right]\left[\begin{matrix}v_x\\ v_y\end{matrix}\right]\nonumber\\
&& +\left[\begin{matrix}N_1(t)\\ N_2(t)\end{matrix}\right]
\end{eqnarray}
which represents a non-linear and a time-varying system. For any finite-bandwidth physical system driven by white noise $W(t)$, the system response function $f(t)$ is assumed to lag the white noise by an infinitesimal time interval \cite{oksen2003}\footnote{This, indeed, forms the basis of It\^o calculus.}. Consequently, $\E[f(t)W(t)]= \E[f(t)|W(t)]\E[W(t)]= 0$. Hence it is easy to see that $\E[N_i(t)]= 0$, $i = 1,2$ because $\theta(t)$ is considered as a response function to $W_x(t)$ and $W_y(t)$. For isotropy we require the variance of fluctuations in the radial velocity component to be independent of $\theta(t)$. It is easy to see that this will be accomplished when $\sigma_{xx} = \sigma_{yy} =: \sigma$ and $\sigma_{yx}=-\sigma_{xy}$. Fig.~\ref{fg:transtraj} shows a sample trajectory in the transverse plane with parameters as shown in the figure. Time is indicated as color on the trajectory. For the parameters chosen, the velocities remain in the range $|v_x|,|v_y|< 13\,$ms$^{-1}$ and show a behavior described by the regime in (\ref{eq:vxs}). It also seen that the position has a drift and is much more correlated in time than the velocity. 
\subsection{Observation Model}
\label{sec:measmod}
In this section we will first present a model that will express the power received in terms of the Cartesian coordinates of the tower $(x_T,y_T,H_T)$ and of the bird $(x,y,z)$. It is known that the Lotek SRX600 receiving unit displays and outputs integer values $Z\in(0,255)$ that are proportional to the power received on a dBm (dBm = 10\,$\log($power in milliwatts)) scale. Keeping this in mind we will then outline a procedure that will determine the constants involved and develop an observation model in terms of the state variables. Finally we will linearize the observation model that will permit application of the extended Kalman filter algorithm.

\subsubsection{Yagi Array Pattern Model} 
\label{sec:Yagi}
\begin{figure}[h!]
\centering
  \begin{subfigure}[htb]{0.475\textwidth}
    \includegraphics[width=\textwidth]{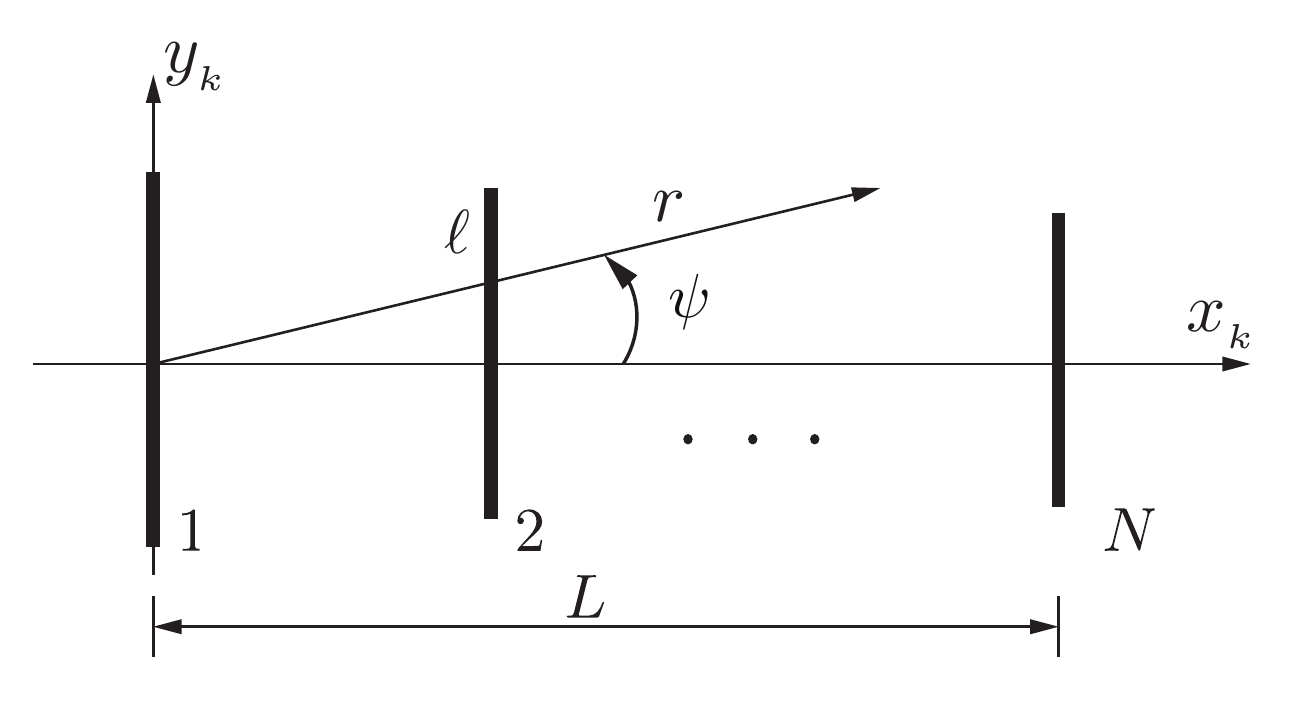}
    \caption{Geometry of the Yagi array.}
    \label{fig:yagi}
  \end{subfigure}
  \begin{subfigure}[htb]{0.475\textwidth}
    \includegraphics[width=\textwidth]{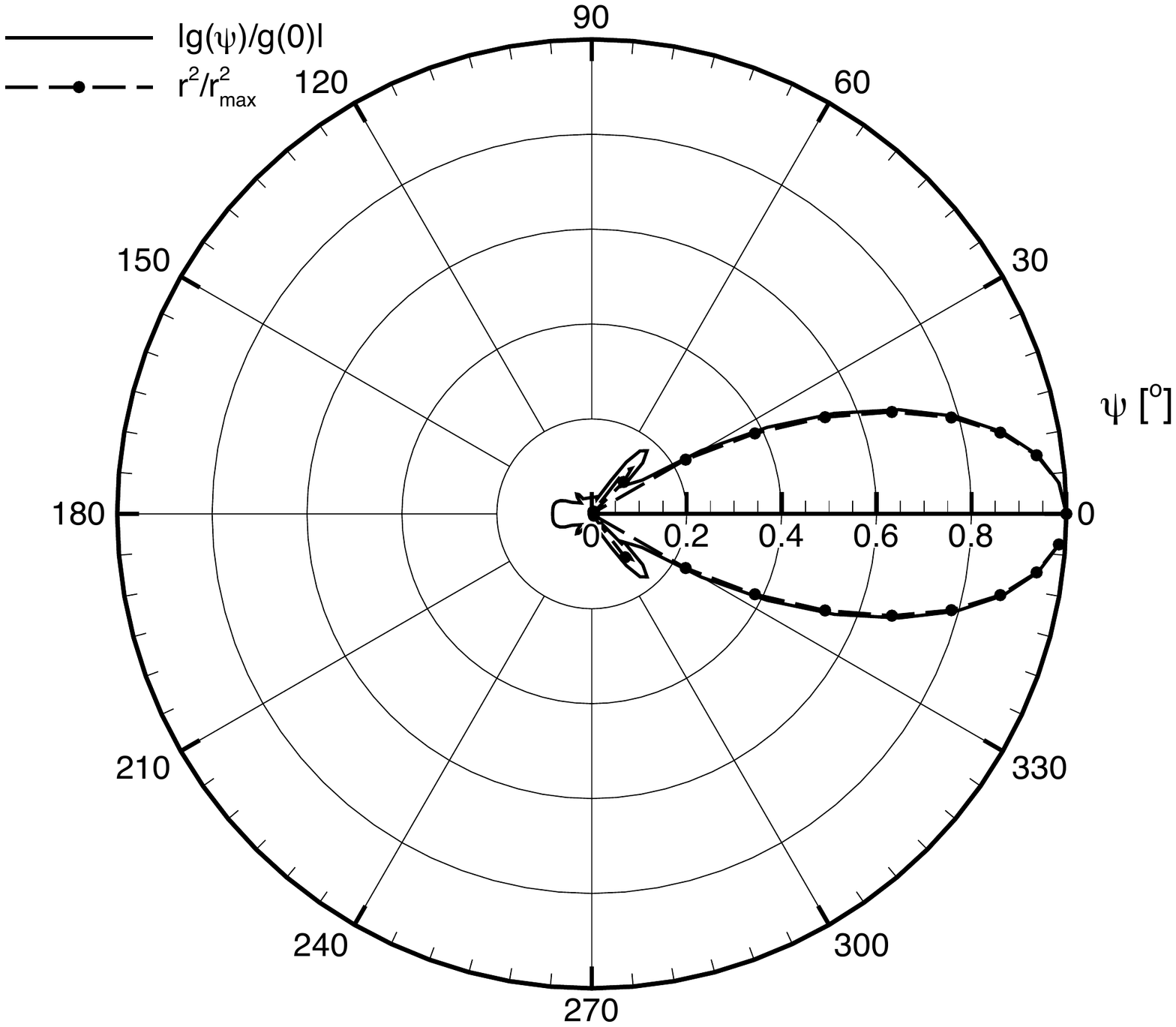}
    \caption{Normalized E-plane pattern (solid line).}
    \label{fig:pattern}
  \end{subfigure}
  \caption{Yagi antenna used in the receiver and its representative pattern in the horizon-\\ tal ($xy$) plane.}
  \label{fg:YagiArray}
\end{figure}

For subsequent analysis, it is required to have knowledge of the radiation pattern of the Yagi antenna. 
Even though the radiation pattern values could be read off from the manufacturer data sheets, it is highly desirable to have an analytical form that will facilitate taking various gradients of the field strength later. The driven and the parasitic elements of the Yagi array all had a length, $\ell$, of around half a wavelength (free-space wavelength $\lambda_0\approx 1.8$\,m) and an overall array length of about 3.5\,m. The array has an absolute gain of around 11.1\,dB (decibels) and a font-to-back ratio of 20\,dB \citep{LTechYagi}. If the main beam of the Yagi points in an azimuthal direction $\phi_T$ in a tower translated local coordinate system and $(r,\phi)$ are the polar coordinates of the point $(x,y)$ relative to the $(x_T,y_T)$: $x-x_T = r\cos\phi, y-y_T = r\sin\phi$, then the angle $\psi$ made by the point $(x,y)$ relative to the main beam is $\psi = \phi-\phi_T$, Figure~\ref{fig:yagi}. The direct line-of-sight distance between the tower and the bird is $R = \sqrt{r^2 + (z-H_T)^2}$. 

For the purpose of developing the observation model, the array factor here is treated as arising from a \emph{continuous} line source of effective length $L_e$ with uniform amplitude distribution and Hansen-Woodyard phase distribution \citep{ST2013}. In the $(x_{k},y_{k})$-plane, the $\psi$-component of the far-zone radiated electric field is given by
\begin{eqnarray}
E_\psi &=& E_0 \frac{e^{-jk_0R}}{R}\frac{\cos\left(\frac{\pi}{2}\sin\psi\right)}{\cos\psi}\frac{\sin(p+q\cos\psi)}{(p+q\cos\psi)}\label{eq:antpat1}\\
&=:&E_0 \frac{e^{-jk_0R}}{R}g(\psi),\label{eq:antpat2}\end{eqnarray}
where $E_0$ is a constant, $j=\sqrt{-1}$, $p = \beta_0L_e/2$, $q=k_0L_e/2$, $k_0=2\pi/\lambda_0$, and the phase per unit length $\beta_0 = -(k_0+2.94/L_e)$ for the Hansen-Woodyard condition. 
Choosing $L_e=4.6\,$m yields a beamwidth of 35.8$^o$ and a front-back ratio of 22\,dB, which are close to the values specified in the data sheets. The normalized field pattern $|g(\psi)|$ versus $\psi$ is shown as solid line in Fig.~\ref{fig:pattern} and closely matches the manufacturer's pattern data despite the increased model length of $L_e = 4.6\,$m versus the actual array physical length of $L=3.5$\,m. For subsequent analysis we write the power gain of the Yagi antenna as $G_0g^2(\psi)$, where $G_0$ is some constant.

In the presence of multipath, equation (\ref{eq:antpat1}) must be modified. If the signal arrives via two paths--one along a direct path between the transmitter and receiver having a path length $R_1 = R$, and an equally strong second one via ground reflection having a path length $R_2 = R+\Delta R$ and reflection coefficient $\Gamma$--then the received electric field is \citep{janas2001}
\begin{equation}
E_\psi \approx E_0 \frac{e^{-jk_0R}}{R}g(\psi)\left[1 + \Gamma e^{jk_0\Delta R}\right].\label{eq:mpath}
\end{equation}
It is assumed in the above equation that both rays arrive in a direction close to the main beam in the elevation plane so that all of the $z$-dependence is contained in $R$ and $\Delta R$ only.   
\subsubsection{Receiver Model}
\label{sec:rxmod}
It was observed experimentally that during certain times the bird was detected via multiple beams and towers that were widely separated. This would be possible if the bird flies at high altitudes where there is substantial height gain associated with large ranges. A height gain will be apparent when there is  (i) a line-of-sight signal between the tower and the bird and (ii) a ground reflected signal arriving from low grazing angles \citep{janas2001} with $\Gamma \approx -1$. The appropriate expression for the received electric field in this situation is (\ref{eq:mpath}) which will result in a height gain factor. If the gain and transmitted power of the radio tag are $P_t$ and  $G_t$, respectively, then the received signal using the two ray model for horizontal polarizations in the absence of noise is \citep{janas2001}
\begin{eqnarray}
P_r &\approx& \frac{P_{t}G_{t}G_0\lambda_0^2g^2(\psi)}{(2\pi R)^2}\sin^2\left(\frac{k_0H_Tz}{R}\right)\nonumber\\
&=:&K_a\frac{g^2(\psi)}{k_0^2R^2}\sin^2\left(\frac{k_0H_Tz}{R}\right)=:K_a\xi^2,\label{eq:Pr}
\end{eqnarray}
where $K_a = P_tG_tG_0$ that depends on the radio tag EIRP (Effective Isotropic Radiated Power $=P_tG_t$)  and the receiving antenna gain factor $G_0$. The factor in (\ref{eq:Pr}) involving the sine function is the height gain factor, which increases monotonically with bird height until the argument becomes equal to $\pi/2$. For most cases encountered in practice the argument of the sine function will be less than $\pi/2$. As an example if $R = 2\,$km, $z = 30\,$m, $H_T = 15\,$m the argument is 0.784. For a given range, the height at which the argument reaches $\pi/2$ is equal to $z_{\rm max} = \lambda_0R/4H_T$. When $R= 1\,$km and receiver height $H_T=30\,$m, $z_{\rm max} =  15\,$m.  

As stated previously, the SRX 600 receiver displays the signal strength, $Z$, using one-byte (i.e., displays integers $Z\in [0, 255]$) with a scale monotonically proportional to the logarithm of the power received. At the low end, the received signal will be corrupted by receiver noise which will cause the display value not go down below certain minimum $Z_m$. At the upper end the receiver is designed to saturate causing the display not to exceed a maximum value, $Z_M$. The range of recorded display values will accordingly fall within the range $0\le Z_m\le Z\le Z_M\le 255$. If 
the minimum receiver power (equal to the receiver noise power) is $P_m = K_aP_0$, the {\em mean} received power in the presence of noise is $P_r = K_a(\xi^2 + P_0)$. 
We model the receiver non-linearity by employing a soft limiter and express the limiter modified measurement, $\cal Z$, as
\begin{eqnarray}
{\cal Z} = \tanh^{-1}\left(\frac{Z-Z_m}{Z_M-Z_m}\right)&\propto& 10\left[\log(P_r)-\log(P_m)\right]\nonumber\\
 &=& 10b_1\log\left(\frac{K_a(\xi^2+P_0)}{K_aP_0}\right)\nonumber\\
 &=:&b\ln\left(\frac{\xi^2}{P_0}+1\right)\label{eq:modeq1}
\end{eqnarray}
where the function $\tanh^{-1}$ is included for soft-limiting, $b_1$ is the constant of proportionality and $b = 10b_1\log(e)$. Equation (\ref{eq:modeq1}) may be rewritten in a simplified form as
\begin{eqnarray}
Z = Z_m + (Z_M-Z_m)\frac{(\xi^2+P_0)^{2b}-P_0^{2b}}{(\xi^2+P_0)^{2b}+P_0^{2b}}\label{eq:xitoZ}
\label{eq:recmod1}\end{eqnarray}
The inverse relation between $\xi$ and $Z$ is easily obtained from (\ref{eq:recmod1}) as
\begin{equation}
\xi^2 = P_0\left[\left(\frac{Z_M-2Z_m+Z}{Z_M-Z}\right)^{\frac{1}{2b}}-1\right].\label{eq:recmod2}
\end{equation}
Given reference measurements $(\xi_i,{\cal Z}_i), i = 1,\dots N$ with known locations of the radio tag,  the constants $b$ and $P_0$ may be estimated by performing non-linear least square fitting. For a given $P_0$ the constant $b$ is equal to
\begin{equation}
b = \frac{\sum\limits_i{\cal Z}_i\ln\left(\frac{\xi_i^2}{P_0}+1\right)}{\sum\limits_i\left[\ln\left(\frac{\xi_i^2}{P_0}+1\right)\right]^2}.
\end{equation}
The constant $P_0$ is then found by minimizing the mean square error between the model prediction and actual data. We conducted reference measurements using a radio tag attached to a kite that was, in turn, attached to a boat whose coordinates were recorded using GPS signals. The boat was made to travel along a straight line in the direction of the main beam of a known tower antenna (MNYN, Fig.\ref{fg:Tower}a). The kite heights were approximately 100\,ft and 200\,ft along the two tracks. Signal from the 100\,ft kite was measured by Southwest beam (beam-5) and that from the 200\,ft kite was measured by the Northwest beam (beam-6).  From the recorder time series signals  we estimate the constants as $b=0.3013, P_0 = 4.8916\times 10^{-11}$. The value for $P_0$ agrees with the estimate $\tilde{P}_0 = kT_0BF/K_a$ based on thermal noise present in the receiver, where $k$ is the Boltzmann's constant, $T_0$ is room temperature, $B$ is receiver bandwidth, and $F$ is its noise figure.  For a radio tag with an EIRP of $0.1\,$mW and a Yagi array with a gain of $G_0 = 10$,  we get $\tilde{P}_0 =  2\times 10^{-11}$ when $T_0 = 290^o\,$K, $B = 1\,$MHz, and $F = 5$. When $\xi^2 = P_0$, equation (\ref{eq:xitoZ}) gives $Z = 0.7941Z_m + 0.2059Z_M$, which for $Z_m = 0, Z_M$ yields $Z = 52.5$. So when the received power equals the noise power generated within the receiver, the display on Lotek SRX600 receiver is predicted to be 52.5. For received signal levels below the noise level, the display values below 52.5 will be unsteady and corrupted by noise.  
\subsubsection{Linearalized Observation Model} 
\label{sec:linmod}
The instantaneous received power in the presence of noise is represented as 
\begin{equation}
P({\bs p},\mu) = (\xi+\sqrt{P}_0\mu)^2=\xi^2+2\xi\sqrt{P}_0\mu+P_0\mu^2,\label{eq:Pinst}
\end{equation}
where $\mu$ is a normal random variable that is statistically independent of the measurement $\xi$. Clearly, the instantaneous power depends on the location coordinates $(x,y,z)$ and noise in a highly non-linear fashion. The mean received power is equal to $\bar{P} = \E(P)=\xi^2+P_0>0$. The variance of the received power is $\tilde{R}(\xi)=\E[(P-\bar{P})^2] = 4\xi^2P_0+2P_0^2$. The mean and the variance are both seen to depend non-linearly on the state vector and it is necessary to linearize the observation equation about a nominal trajectory $({\bs p},\mu) = (\tilde{\bm p},0)$ before applying the standard Kalman filtering\footnote{The standard Kalman filter algorithm applies when both the system model and the observation model are linear in the state-space vector $\bm{p}$. However, when either one of the system or observation models is nonlinear in $\bm{p}$, the model must first be linearized about a nominal trajectory and then the linear algorithm can be applied. This is known as the extended Kalman filter.} technique. Recall that
\begin{align}\label{eq:rely}
r^2&=(x-x_T)^2+(y-y_T)^2&R^2&=r^2+(z-H_T)^2\\
\psi &= \phi-\phi_T&\tan\phi &= \frac{y-y_T}{x-x_T}\\
\frac{\partial \psi}{\partial y} &=\frac{x-x_T}{r^2}, \ \frac{\partial}{\partial x_z} = 2x_z\frac{\partial}{\partial z}& \frac{\partial \psi}{\partial x} &=-\frac{y-y_T}{r^2}
\end{align}

Expanding $P$ in a Taylor's series about a nominal trajectory $(\tilde{\bs p},0)$ and keeping only the linear terms in $\bs p$ and noise (which is now labeled as $\zeta$) yields
\begin{equation}	
P(\bs{p},\zeta)  \approx  {\bf H}(\tilde{\bs p}){\bs p} +d +\zeta =:Y({\bs p})\label{eq:measmod}
\end{equation}
where $d = h(\tilde{\bs p}) - {\bf H}(\tilde{\bs p}) \tilde{\bs p}$, $h(\tilde{\bs p}) = \tilde{\xi}^2 + P_0$, $\tilde{\xi} = \xi(\tilde{\bs p})$, $\zeta = \sqrt{\tilde{R}(\tilde{\xi})}\,{\cal N}(0,1)$, ${\cal N}(0,1)$ being the normal random variable, ${\bf H}(\tilde{\bm p})$ is the $1\times 5$ measurement row vector equal to 
\begin{equation}
{\bf H}(\tilde{\bm p}) = 2\tilde{\xi}\left(\frac{\partial \xi}{\partial x},\ 0,\ \frac{\partial \xi}{\partial y},\ 0, \frac{\partial \xi}{\partial x_z}\right)\bigg\vert_{\tilde{\bm p}}.
\end{equation}
\subsection{Location Tracking by Kalman Filtering}
\label{sec:Kal}
Given the system model, (\ref{eq:sol}), the observation model (\ref{eq:measmod}), a set of measurements $Y_i=Y[\bm{p}(t_i)], i=1,2,\ldots$, system and observation noise covariance parameters $\sigma_{xx}$, $\sigma_{yy}$, $\sigma_{zz}, \sigma_{xy}, \sigma_{yx}, \sigma_{zx}, \sigma_{zy}$ and $P_0$, the {\em conditional mean\/} of bird location can be estimated by using the extended Kalman filtering technique as it evolves in time. What are recorded by the measurement system are the corrupted (by measurement noise) signal powers associated with a particular bird trajectory. The estimated mean bird trajectory will be dictated both by the system model and the measurements available. As a byproduct the variances associated with the mean trajectory are also calculated by the Kalman filter. For LTI systems operating in the presence of Gaussian noise, the Kalman filter gives the best estimate based on the current measurement. The following notation is first introduced:
\begin{eqnarray}
\hat{\bm{p}}_i^-&=&\mE[\bm{p}_i|Y_1,Y_2,\ldots,Y_{i-1}], \ \mbox{\emph{a priori} estimate of } \bm{p}_i,\notag\\
\hat{\bm{p}}_i^+&=&\mE[\bm{p}_i|Y_1,Y_2,\ldots,Y_i],\ \mbox{\emph{a posteriori} estimate of }\bm{p}_i\notag\\
{\bs P}_i^-&=&\mE[(\bm{p}_i-\hat{\bm{p}}_i^-)(\bm{p}_i-\hat{\bm{p}}_i^-)']\notag\\
&&(\mbox{covariance of estimation error of }\hat{\bm{p}}_i^-)\nonumber\\
{\bs P}_i^+&=&\mE[(\bm{p}_i-\hat{\bm{p}}_i^+)(\bm{p}_i-\hat{\bm{p}}_i^+)']\notag\\
&&(\mbox{covariance of estimation error of }\hat{\bm{p}}_i^+)\nonumber\\
R_i&=&\mbox{covariance of measurement error}\notag\\
\hat{\bm{p}}_0^+&=&\mE[\bm{p_0}],\quad
{\bs P}_0^+=\mE[(\bm{p}_0-\hat{\bm{p}}_0^+)(\bm{p}_0-\hat{\bm{p}}_0^+)']\notag\end{eqnarray}
The nominal trajectory for linearizing the measurement equation is chosen to be $\tilde{\bm{p}}=\hat{\bm{p}}_i^-$ so that ${\bf H}_i:={\bf H}(\hat{\bm{p}}_i^-)$, $d_i := h(\hat{\bm{p}}_i^-) - {\bf H}_i\hat{\bm{p}}_i^-$. The discrete-time extended Kalman filter, \citep{DSimon2006}, \citep{DK2012} is given by the following equations, which are computed for each time step $i=1,2,\ldots$:
\begin{eqnarray}
 \hat{\bm{p}}_i^-&=&{\bf T}_{i-1}\hat{\bm{p}}_{i-1}^+\label{eq:pred}\\
{v}_i &=&Y_i-{\bf H}_i\hat{\bs p}_i^- - d_i = Y_i - h(\hat{\bs p}_i^-)\\
{\bs P}_i^-&=&{\bf T}_{i-1}{\bs P}_{i-1}^+{\bf T}_{i-1}'+{\bs Q}_{i-1}\\
 F_i&=&{\bf H}_i{\bs P}_i^-{\bf H}_i' + R_i\\
 {\bs k}_i&=&{\bs P}_i^-{\bf H}_i'F_i^{-1}\\
 \hat{\bm{p}}_i^+&=&\hat{\bm{p}}_i^-+{\bs k}_i{v}_i\label{eq:corr}\\
 {\bs P}_i^+&=&({\bs I}-{\bs k}_i{\bf H}_i){\bs P}_i^-.
 \end{eqnarray}
 
It is important to note that when the measurement noise is very high, $F_i\to R_i$ and the norm of the Kalman gain vector $\Vert{\bs k}_i\Vert\to 0$. Consequently, $\hat{\bs p}_i^+\to\hat{\bs p}_i^-$, meaning that the mean trajectory is governed primarily by the `predictor' part of the Kalman filter or, equivalently, by the noise-free system model (\ref{eq:pred}). On the other hand, when the system noise is very high, ${\bs P}_i^-\to{\bs Q}_{i-1}$, $F_i\to{\bf H}_i{\bs Q}_{i-1}{\bf H}_i'$ and the `corrector' part of the Kalman filter (\ref{eq:corr}) will also gain prominence. 
 
For the Kalman filter to estimate the states effectively, the system must be {\em observable\/}. Observability depends on the system coefficient matrix ${\bf T}_i$ and the measurement matrix ${\bf H}_i$. For an LTI system to be {\em completely observable}, the $5\times 5$ observability matrix ${\bs S} = [{\bf H}_i, {\bf H}_i{\bf T}_i, {\bf H}_i{\bf T}_i^2, {\bf H}_i{\bf T}_i^3, {\bf H}_i{\bf T}_i^4]'$ must be full rank \citep{ham1983}. However, given that the system models along the three spatial coordinates are uncoupled and power measurements are independent of the individual's speed, the observability matrix can at best have a rank of 4 in our case. So that the system is {\em almost observable} (with a degree of observability of $4$) we have to choose the system parameters such that rank of $\bs S$ does not fall below 4. A necessary condition for the rank to be at least 4 with $|\partial\xi/\partial x|, |\partial\xi/\partial y|, |\partial\xi/\partial x_z|\ne 0$ is that $\beta_x\ne\beta_y\ne \beta_z$. Hence isotropy must be sacrificed in order to meet the observability condition. We choose $\beta_x\approx\beta_y\ne\beta_z$ to meet the almost observability condition.

The more complete dynamic prediction that is achieved through the proposed dynamical model and Kalman filtering may be contrasted with the static prediction of the triplet $(r,\psi,z)$ by directly inverting equation (\ref{eq:Pr}) without recourse to any dynamic system. For a given value of measured signal $Z$, we have from (\ref{eq:Pr}) in a noise-free environment that
\begin{equation}
\frac{(k_0R)^2}{\sin^2\left(\frac{k_0H_Tz}{R}\right)} = \frac{g^2(\psi)}{\xi^2}.\label{eq:stat1}
\end{equation}
It is apparent from (\ref{eq:stat1}) that a multitude of $(r,\psi,z)$ combinations can give rise to the same value of $\xi$ (and hence $Z$) as we have observed in $Ob_1$. For bird and tower heights much smaller than the horizontal bird range $r$ to tower, $k_0H_Tz\ll R$, we have the approximation $R^2 = (x-x_T)^2+(y-y_T)^2+(z-H_T)^2\approx (x-x_T)^2+(y-y_T)^2=r^2$ and $\sin(k_0H_Tz/R)\approx k_0H_Tz/r$ in (\ref{eq:stat1}), thus resulting in 
\begin{equation}
r^2 \approx \frac{H_Tz}{|\xi|}|g(\psi)|.\label{eq:stat0}
\end{equation}
Hence the range squared determined for any given measurement follows the shape of the field radiation pattern of the Yagi array. The dot-dashed line in Fig.~\ref{fig:pattern} shows $r^2$ normalized to its maximum possible value, $r_{\rm max}^2$ for a given measurement $|\xi|$ and bird height $z$. Thus a given measurement value could arise from a nano-tag located at large ranges within the 3\,dB beam width of the array or from relatively shorter ranges located along the side lobes or back lobe of the array. Note also that the range scales as $\sqrt{|g(\psi)|}$. So the same signal can be received from a bird that is, say, at 6\,km along the main beam or at 1.1\,km along a $-15\,$dB sidelobe or at 600\,m along a $-20\,$dB back lobe. These numbers are representative of our Yagi array whose pattern is shown in Fig.~\ref{fig:pattern}. However, if the height $z$ and bearing $\phi$ of the bird are {\em both} known by some other means, then equation (\ref{eq:stat0}) can be used to {\em uniquely} determine the range to a measuring tower. Equation (\ref{eq:stat0}) also shows that for a given measurement $Z$, longer ranges to the tower are possible if the bird flies at higher altitudes as we have indicated in $Ob_3$.

Under the assumption that the signals arrive {\em primarily\/} through the main beam of the receive array so that $g(\psi)\approx g(0)$, we get 
\begin{equation}
\frac{r^2}{z}=\frac{H_T|g(0)|}{|\xi|}=\frac{H_T}{|\xi|}\frac{\sin(p+q)}{(p+q)},\quad k_0H_Tz\ll r,\  |\psi| < \Theta,\label{eq:stat2}
\end{equation}
where $\Theta$ is the half-power angle of the receiving antenna. (For the Yagi antenna used in our measurements $\sin(p+q)/(p+q) = 0.6768$.).  Equation (\ref{eq:stat2}) constitutes what we label as the {\em static model}. 
\section{Results}\label{sc:res}
\subsection{Validation Using Simulated Bird Trajectory}
\label{sec:simul}
\begin{figure}[h!]
\centering
  \begin{subfigure}[htb]{0.475\textwidth}
    \includegraphics[width=\textwidth]{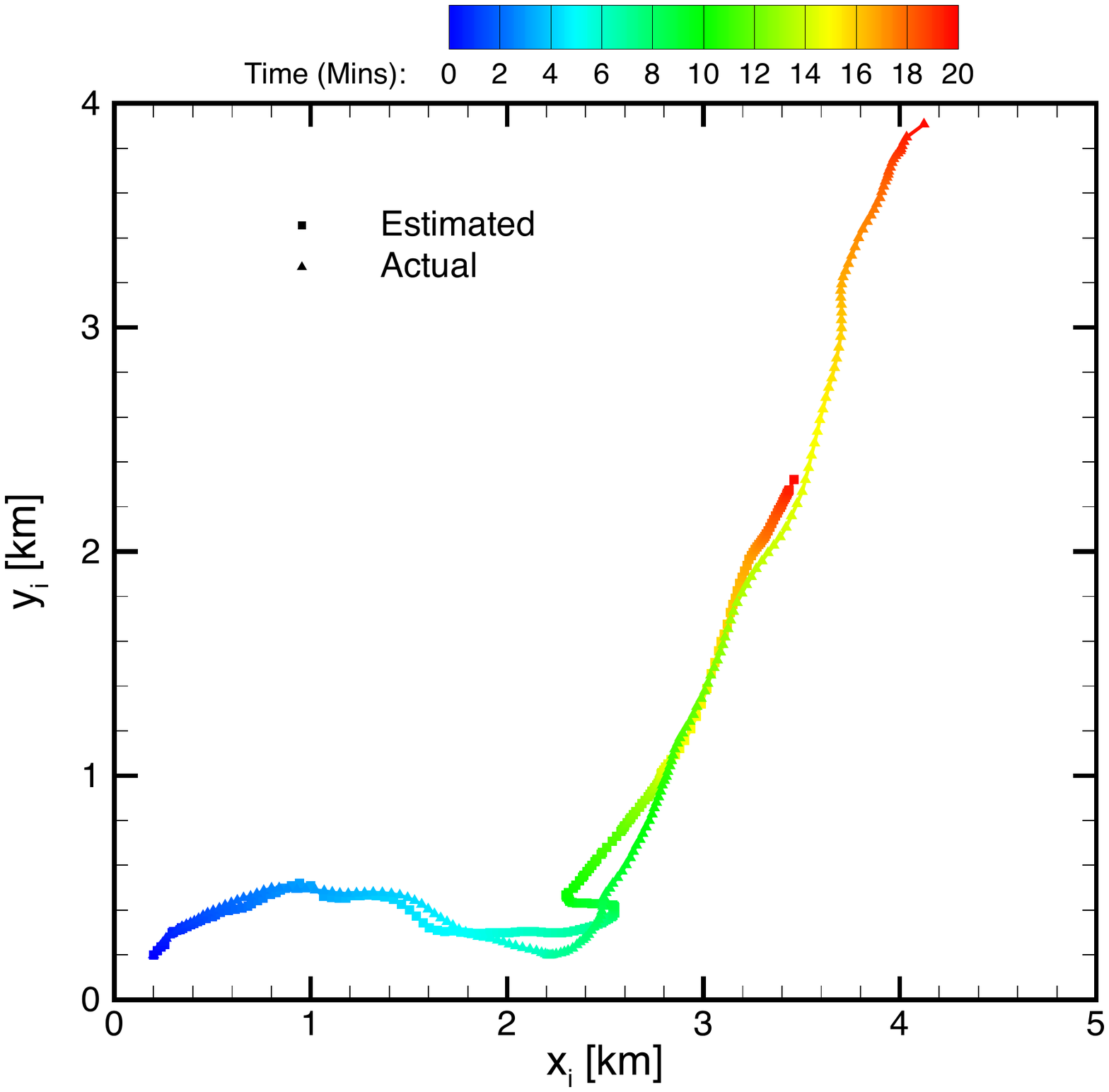}
    \caption{Transverse Plane.}
    \label{fg:PosMock}
  \end{subfigure}
  \begin{subfigure}[htb]{0.475\textwidth}
    \includegraphics[width=\textwidth]{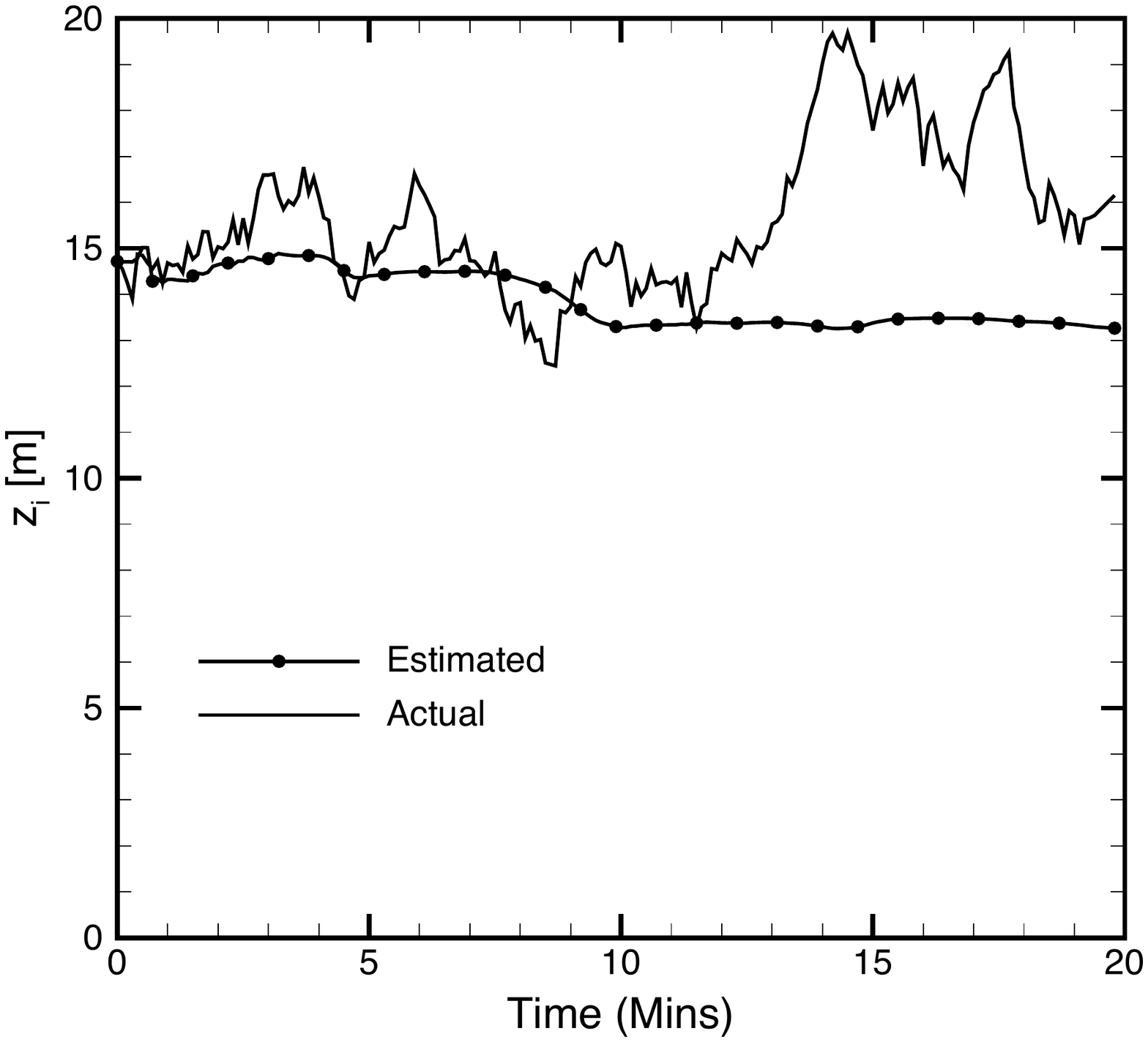}
    \caption{Vertical Direction.}
    \label{fg:Mockht}
  \end{subfigure}
  \caption{Comparison between estimated and actual locations for a simulated traject-\\ ory. Distances shown are relative to tower coordinates $(x_T, y_T, 0)$. The various paramet-\\ ers used in the simulation were $\beta_x = 2.5\times 10^{-4}, \beta_y = 2.25\times 10^{-4}, \beta_z = 1\times 10^{-5}$,\\ $\sigma_{xx} = 0.25=\sigma_{yy}, \sigma_{zz} = 0.02, \sigma_{xy}=0.25\sigma_{xx}=-\sigma_{yx}, \sigma_{zx} = 0.5\sigma_{zy} = 0.2\sigma_{zz}$.} 
  \label{fg:Simtraj}
\end{figure}

\begin{figure}[h!]
\centerline{\scalebox{0.375}{\includegraphics{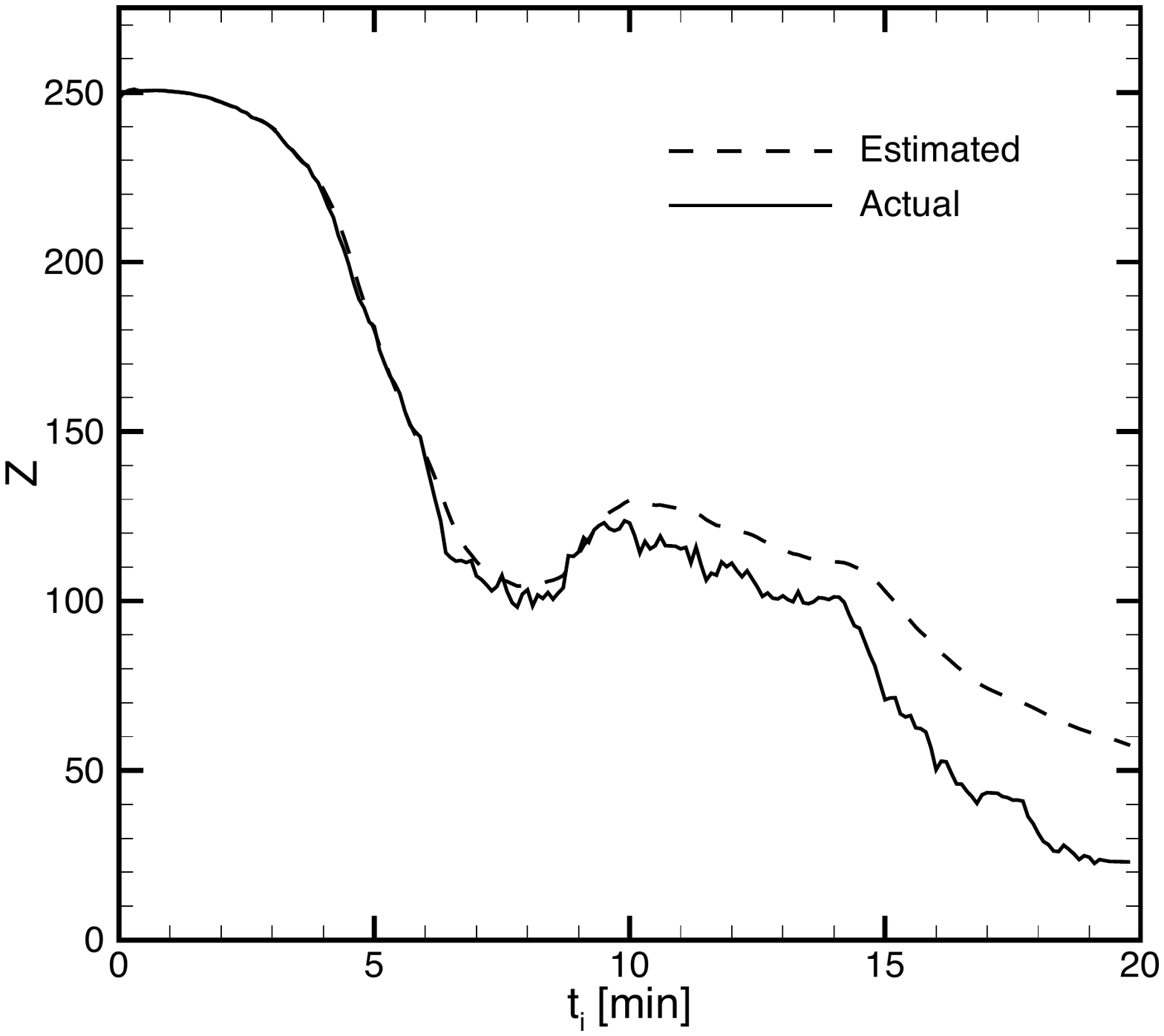}}}
\caption{Calculated display values of received signal strength $Z$ for the actual- and \\ estimated trajectories.}
\label{fg:ZData}
\end{figure}

Assuming tower coordinates of $x_T=417768$ and $y_T = 4606808$ (coordinate system UTM Zone 19N, units in m) we generated a simulated trajectory using equations (\ref{eq:sol}) for a bird starting at $x = x_T+200, y = y_T+200$ with velocity components $v_x(0) = 2\sqrt{2}$, $v_y(0) = 2\sqrt{2}$ and an initial height of $z (0) = 14.72\,$m (equal to the tower height above sea level). Locations were simulated every 6 seconds (equivalent to the burst rate interval of the nano-tag transmitter) for a total of 1,200 seconds. Then with the assumed tower coordinates and choosing the appropriate beam, we calculate the horizontal distance, height and bearing. The corresponding values of $Z$ and $Y$ are calculated using (\ref{eq:Pr}), (\ref{eq:modeq1}) and (\ref{eq:recmod1}). In the Kalman filter estimation, we have excluded points that produce the display values $Z<22$ which correspond to received signal-to-noise ratio of less than $-4.7\,$dB. The selected values are then input to the Kalman filter and state vector estimated for each instant of time. Fig~\ref{fg:Simtraj} shows the comparison between the estimated trajectory and the actual trajectory in both planes. It is to be noted that the conditional mean trajectory estimated by the Kalman filter is not expected to exactly follow the actual trajectory, which is all but one realization of a whole ensemble of possible trajectories. However, the predicted trajectory is seen to capture the long term trends quite well including turnings in the horizontal plane even though the range at end of the interval falls short by about 1.6\,km. For the predicted trajectory we calculate the display values as computed from (\ref{eq:recmod1}) and compare them to the actual Z-values in Fig.\ref{fg:ZData}. It is again seen that the display values follow the trends very well. 

\subsection{Estimated Trajectory for a Migrating Piping Plover}\label{sec:estimated}
\label{sec:mig}
We next apply the algorithm to predict the location of an adult Piping Plover (Charadrius melodus) that was tracked using multiple beams/towers during its migratory departure from the breeding grounds  Fig.~\ref{fig:PamFig1}. The Piping Plover was captured during the incubation period using a walk-in trap \citep{Bub1991} at its nest on Aquidneck Island, Rhode Island, USA. The plover was fitted with a nano-tag that was attached to its inter-scapular region using epoxy. Signals from the bird arrived at the towers at highly irregular intervals interspersed with several large gaps. We will show the predictions carried out over one large subinterval during which the bird was thought migrating in a southwest direction. The radio tag on the bird was detected by several towers that included SACH, TRUS, NAPA, BISE, PLIS, and MNTK (see Fig.~\ref{fig:PamFig1}). Several hours later it was detected in New Jersey by a tracking station on the Motus network (www.motus.org). As remarked earlier, numerous trajectories can all lead to the same set of measurements. Because tracking is recursive in nature and is intimately tied to the initial state, some strategy must be employed to narrow down on the number of possible trajectories. Different strategies will lead to different estimated trajectories because of the non-uniqueness of the problem at hand. In the results we show, we have employed the following assumptions and strategies:
\begin{enumerate}
\item An initial height $z_0$ is {\em chosen} based on previous or expert knowledge.  
\item The system parameters $\beta_x, \beta_y, \beta_z,  \sigma_{xx}=\sigma_{yy}, \sigma_{xy} = -\sigma_{yx}, \sigma_{zz}, \sigma_{zx} = \sigma_{zy}$ are {\em chosen} based on a trail and error basis. From isotropy and observability criteria $\beta_y\simeq\beta_x$. Our numerical trials have indicated that a good rule of thumb for modeling migratory type of behavior is $10^{-5}[{\rm s}^{-1}]\le \beta_x,\beta_y\le 10^{-3}[{\rm s}^{-1}]$. From (\ref{eq:vxs}) it is seen that the condition $\beta_x\Di\ll 1$ is satisfied as long as the maximum time gap between successive data arrivals at an antenna satisfies $\Di\le 100$\,s. For a reasonable random change in speed of about 1\,m/s between successive data observations in the transverse plane, the noise parameters may be chosen according to $\sqrt{a\Di}\simeq 1$ following (\ref{eq:vxs}). For example with $\sigma_{xy}=0.5\sigma_{xx}$, $\Di = 80\,$s, one gets the values $\sigma_{xx}\simeq 0.1\,$[ms$^{-3/2}$] and $\sigma_{xy} = 0.05$\,[ms$^{-3/2}$]. Similar order of values for $\beta_z, \sigma_{zz}$, and $\sigma_{zx}$ also implied from (\ref{eq:xz}). \label{itm:par}
\item A maximum possible radial speed $v_{\rm max}$ for the bird is {\em specified}.
\item An initial covariance matrix ${\bs P}_0^+$ is {\em chosen}. A reasonable choice is to choose ${\bf P}_0$ to be diagonal with entries proportional to the uncertainties of the state vector. For example, we chose $P_0(1,1) = 10,  P_0(2,2) = 10, P_0(3,3) = 10,  P_0(4,4) = 10, P_0(5,5) = 100$ here.
\item The initial state is determined by considering the first two measurement points $\xi_0,\xi_1$ recorded at times $t_0,t_0+\Delta t$, respectively. The sets of all possible $(x_0, y_0, z_0), (x_1,y_1,z_0)$ are determined by inverting the non-linear equation (\ref{eq:stat1}) with $\xi=\xi_0, \xi_1$, respectively. We narrow the sets by imposing the constraint $\sqrt{(x_1-x_0)^2 + (y_1-y_0)^2}/\Delta t
\le v_{\rm max}$ on the maximum radial speed. From the narrowed sets we pick the pair $(x_0,y_0,z_0), (x_1,y_1,z_0)$ and take the initial state vector as $X_0 = [x_0, (x_1-x_0)/\Delta t, y_0, (y_1-y_0)/\Delta t, \sqrt{z_0}]'$. We run the complete Kalman filter over the time interval $(0,T)$ for each initial state in the narrowed sets. For the trajectory so produced, we determine the estimated received signal $\hat{Z}$ using (\ref{eq:Pr}) and (\ref{eq:recmod1}) and calculate the root mean square error 
\begin{equation}
\delta Z = \sqrt{\frac{1}{T}\int\limits_0^T\left(\hat{Z}-Z\right)^2\,dt}
\end{equation} We pick the initial state that minimizes this error. Alternatively, one could have minimized the integral of the norm of the covariance matrix ${\bs P}_i^+$. \label{itm:init}
\item When the duration, $T_g$, between two successive measurements exceeds a certain chosen maximum gap time we rerun the prediction all over starting at the end of that time gap.  If $(x_p,y_p,z_p)$ are the location coordinates of the bird just prior to the gap, the possible initial location pairs $(x_0,y_0)$ and $(x_1,y_1)$ are determined as in item~\ref{itm:init} for $z_0 = z_p$. We impose the additional constraint $R=\sqrt{(x_0-x_p)^2 + (y_0-y_p)^2} \le v_{\rm max}T_g$. We then choose the $(x_0,y_0)$ that corresponds to the maximum $R$ to avoid stagnation and pick the corresponding initial speeds.  \label{itm:gap}
\item In the rare instance of simultaneous recordings at two different towers/beams, the state vector is determined separately for each tower/beam and the final state vector is taken as the average of the two.
\end{enumerate}

%Fig~\ref{fg:Subset1} shows the estimated trajectory when the bird (piping plover) was tracked by Sachuest tower (SACH in Fig.~\ref{fig:PamFig1}) and was likely hovering near its nest and making localized movements at low altitudes. The reference coordinates correspond to the base coordinates of Sachuest tower. Data was received at highly irregular intervals and contained four large time gaps that exceeded 120\,s. Gap treatment was done as described in item \ref{itm:gap} above. The transverse and vertical coordinates shown for the various clusters in Fig.~\ref{fg:Subset1} confirm that the bird was in a nesting type of mode during the 90 minute tracking time.  Dotted lines connecting the clusters are included for visual clarity. 

\begin{figure}[h!]
\centering
  \begin{subfigure}[htb]{0.475\textwidth}
    \includegraphics[width=\textwidth]{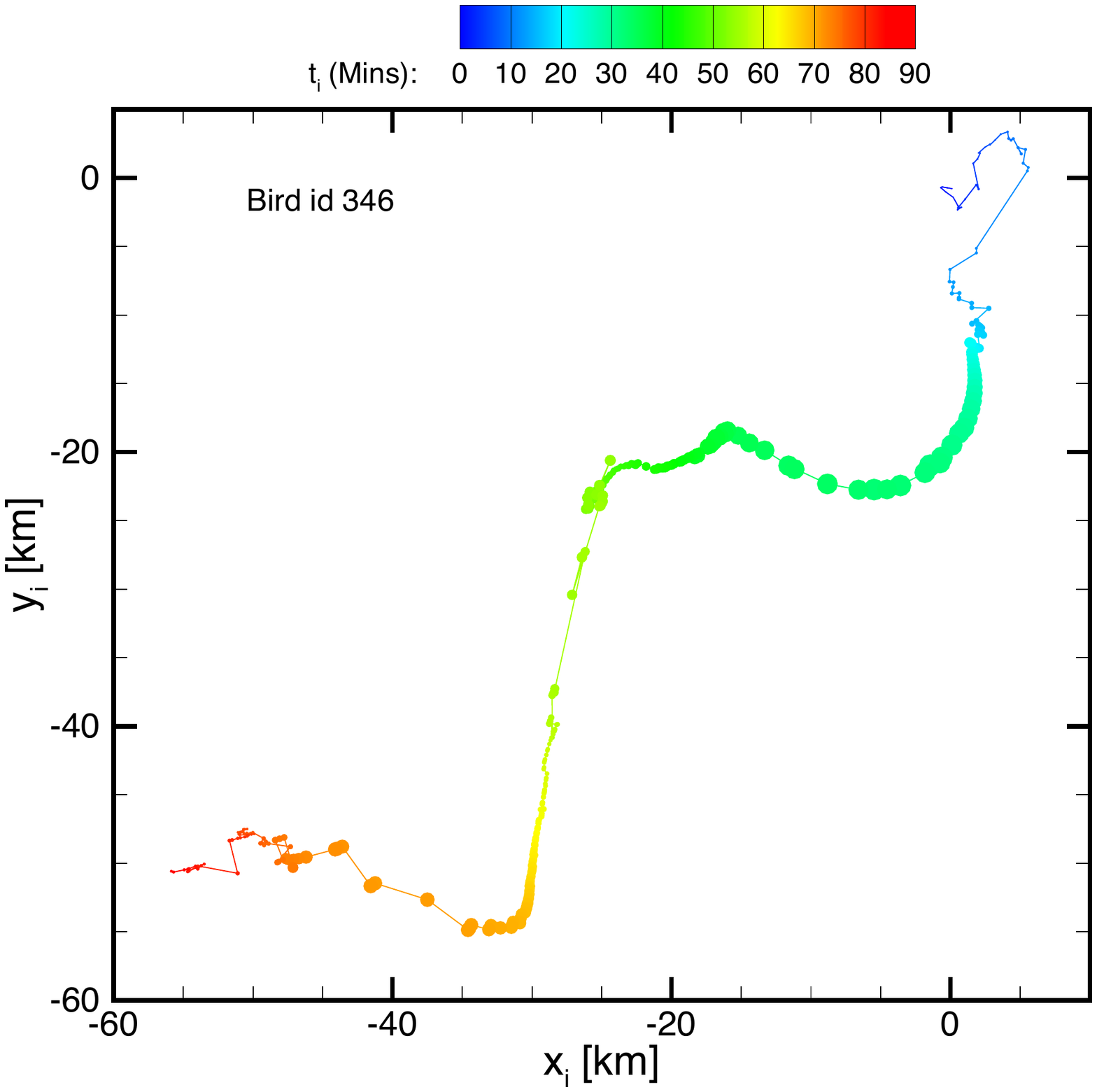}
    \caption{Transverse Plane.}
    \label{fig:xy2}
  \end{subfigure}
  \begin{subfigure}[htb]{0.475\textwidth}
    \includegraphics[width=\textwidth]{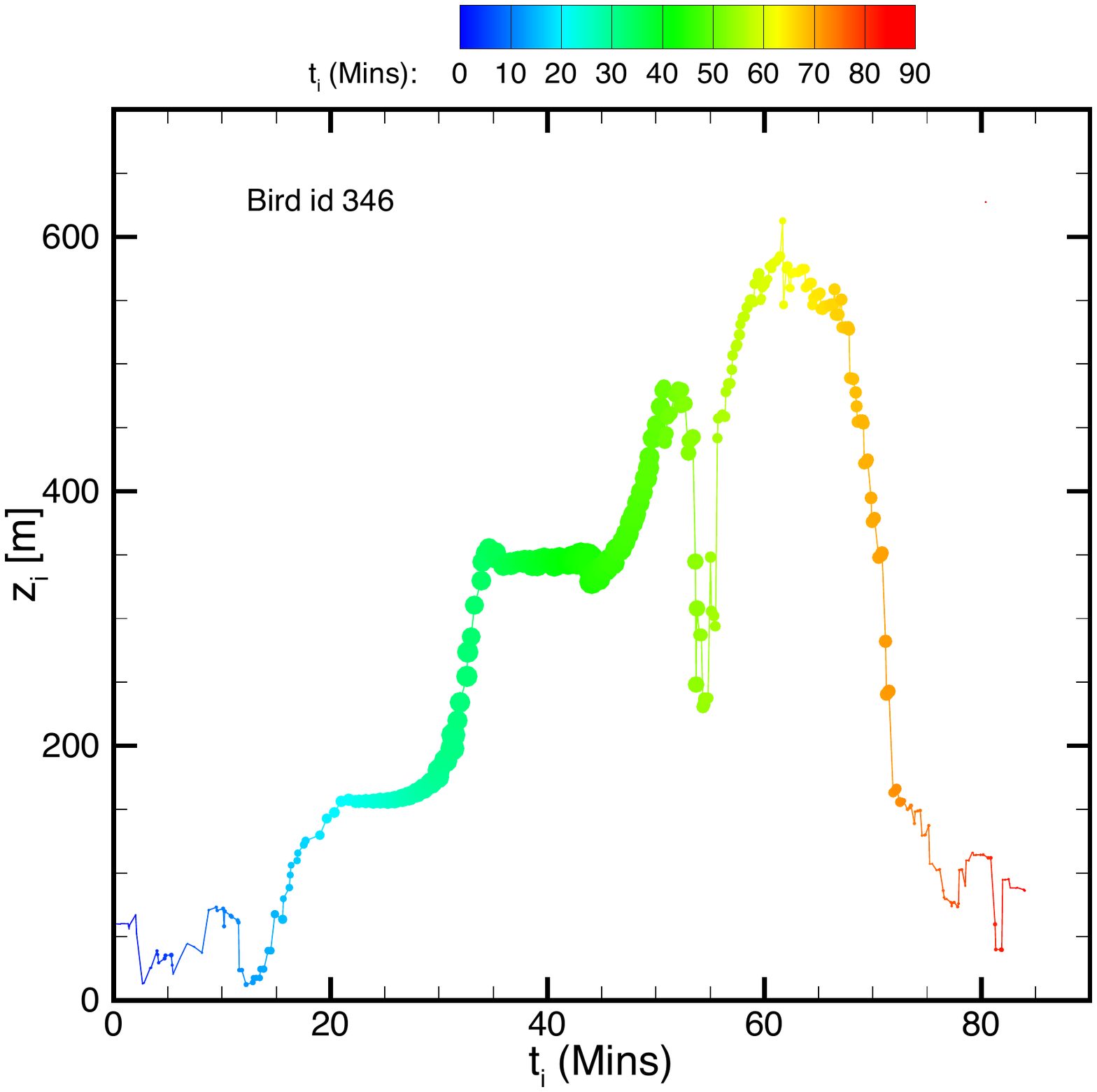}
    \caption{Vertical Direction.}
    \label{fig:z2}
  \end{subfigure}
  \caption{Estimated coordinates of a migrating Piping Plover tracked by multiple\\ towers during a sub-interval. Parameters used were $\beta_x = 2.5\times 10^{-4}, \beta_y = 2.25\times 10^{-4},\\ \beta_z = 1\times 10^{-5},\sigma_{xx} = 0.3 = \sigma_{yy}, \sigma_{zz} = 0.125, \sigma_{xy} = 0.5\sigma_{xx}, \sigma_{xz} = 0.5\sigma_{zz}, z_0 = 60\,$m.\\ The transverse coordinates are relative to the base coordinates of Sachuest tower. \\ Symbol size is equal to two tenths of actual standard deviation of error in range and \\ one half of actual standard deviation of error in height. Maximum (median) standard\\ deviation in range is 8\,km (2.7\,km) and maximum (median) standard deviation in height\\ is 41.5\,m (17.3\,m).} 
  \label{fg:Subset2}
\end{figure}

Fig.~\ref{fg:Subset2} shows the estimated bird trajectory. System parameters and initial conditions used in the algorithm are shown in figure caption and inset. 

\begin{figure}[h!]
\centerline{\scalebox{0.375}{\includegraphics{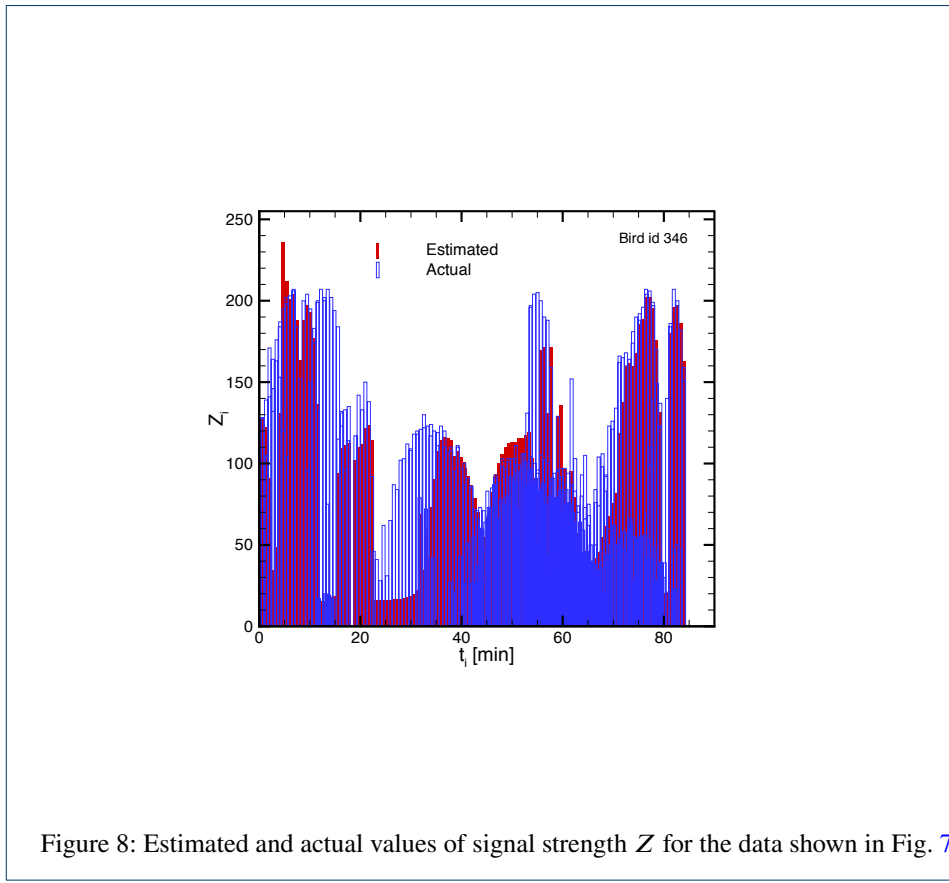}}}
\caption{Estimated and actual values of signal strength $Z$ for the data shown in Fig.~\ref{fg:Subset2}.}
\label{fg:Zbcomp}
\end{figure}

It is evident that the overall model is able to predict a 3-dimensional flight trajectory spanning $>$ 50\,km and associated flight altitudes (spanning hundreds of meters). The somewhat large spike seen in the bird altitude at $t = 54\,$minutes is likely caused by erroneous recorded data coupled with a relatively large time gap. Large time gaps in data also cause relative large standard deviations of error in the estimated locations. In Fig.~\ref{fg:Subset2}, the median (maximum) standard deviation of error in range estimated by the filter is about 2.7\,km (8\,km) and the median (maximum) standard deviation of error in height is about 17.3\,m (42\,m). In general higher error is resulted when the time gap in the data exceeds a preset threshold and the filter has to reinitialize. The fidelity of estimated trajectory is measured by comparing the recorded signal values $Z$ with the estimated values $\hat{Z}$. Fig.~\ref{fg:Zbcomp} shows the comparison over the 90 minute second subinterval. It is seen that the estimated $\hat{Z}$ tracks the recorded $Z$ well over a large dynamic range, thus giving credibility to the estimation. 

\section{Discussion}
\label{sec:disc}
From a computational perspective, the technique presented is remarkable in the sense that four simultaneous detections are normally needed with omnidirectional antennas to determine the 3D position of an object uniquely, whereas the technique presented allows plausible solutions based on a single power measurement only. However, since any given signal strength from a 3D Yagi beam pattern can result from a broad range of locations, inversion to location is inherently non-unique, {\em i.e.}, the system is non-observable in the strictest sense. In order to obtain unique solutions additional constraints have to be imposed during the prediction process. In this study, we imposed a constraint on the maximum radial speed and employ an error minimization procedure, but chose the system parameters on a trial and error basis. Guidelines for choosing the system parameters for migratory behavior are provided in \S\ref{sec:mig}, item \ref{itm:par}. A more systematic approach might be to base the choice on a parameter estimation technique with maximum likelihood \citep{DK2012}; however, our initial attempts at doing so resulted in optimized parameters that did not differ substantially from those presented here. More detailed investigations should be done in this direction and will be explored in the future.\par

Further research is also needed to quantify the model's ubiquity, precision and robustness to the quantity of measurements. For example, while accounting for detection by side lobes and the back lobe of antenna enhances the potential accuracy and precision of the technique, the model did exhibit some sensitivity to uncertainty in a bird's initial location, transverse speeds and initial error covariance matrix as well as to the system parameters (which in our case reduced from ten to eight using isotropy considerations). Another challenge occurs when the available data contains many large time gaps, which would necessitate special treatment of the state vector at the gaps. The standard deviation of estimated error will depend on how the gap is treated. One possible scheme of gap treatment is described in \S\ref{sec:estimated}, item \ref{itm:gap}. However, other meaningful strategies will also be explored in the future.

\section{Conclusions}
\label{sec:concl}
The ability to generate three-dimensional flight paths of small-bodied airborne organisms across extended spatio-temporal domains can help both answer fundamental ecological questions and address conservation and management concerns. We have described a new technique based on state-space models and Kalman filtering to predict the dynamic 3D trajectory of a radio-tagged bird given the animal's initial state. This represents the first such model to incorporate radio multipath phenomenon, allowing for an improved estimate of range, bearing, and altitude of radio-tagged animals according to the radiation characteristics of the sensing Yagi array. Soft limiting in the received signal has been included to capture the true input-output behavior at low and high signal levels. Incorporation of Cox-Ingersoll-Ross type movement model in the vertical plane ensures that the bird's altitude remains above sea level at all times, and realistic trajectories were facilitated through the use of an Ornstein-Uhlenbeck type movement model in the transverse plane, with all three spatial coordinates being statistically correlated.\par

From this applied ecology perspective, we feel that the technique is flexible and broadly applicable to other telemetry networks and ecological systems. We encourage researchers both to adopt the correlation motion model in three dimensions when designing state-space models and to compare existing triangulation methods \citep{Smolinsky2013} and antenna pattern models \cite{mitchell2015, Smetzer2017} with our novel 3D receiver model and its simplifications (\ref{eq:stat1})-(\ref{eq:stat2})). Thus, given appropriate array configuration and modification of the observation and process models, our technique could be applied to the analysis of automated VHF telemetry data across a wide range of spatial and temporal scales, from site-specific studies using targeted arrays, to coordinated digital VHF tracking efforts that span the Hemisphere.  
%\nocite{oreg,schn,pond,smith,marg,hunn,advi,koha,mouse}

%%%%%%%%%%%%%%%%%%%%%%%%%%%%%%%%%%%%%%%%%%%%%%
%%                                          %%
%% Backmatter begins here                   %%
%%                                          %%
%%%%%%%%%%%%%%%%%%%%%%%%%%%%%%%%%%%%%%%%%%%%%%
\begin{backmatter}
\section*{List of Abbreviations}
3D = Three-dimensional\\
dB = decibels\\
EIRP = Effective Isotropic Radiated Power\\
GPS = U.S. Global Positioning System\\
ID = Identity\\
LTI = Linear-Time-Invariant\\
OU = Ornstein-Uhlenbeck\\
U.S. = United States\\
VHF = Very High Frequency (30 MHz to 300 MHz)
\section*{Declarations}
\subsection*{\em Ethics Approval and Consent to Participate}
Placement of radio tags on birds was approved by the US Department of the Interior, Fish and Wildlife Service.
\subsection*{\em Competing Interests}
  The authors declare that they have no competing interests.
\subsection*{\em Funding}
This study was funded in part by the U.S. Department of the Interior, Bureau of Ocean Energy Management through Interagency Agreement M13PG00012 with the U.S. Department of the Interior, Fish and Wildlife Service.  This study was also supported through the NSF-sponsored IGERT: Offshore Wind Energy Engineering, Environmental Science, and Policy (Grant Number 1068864). 
\subsection*{\em Author's contributions}
RJ developed the theory, the system and the observation models. PL installed the telemetry towers and carried out the measurements. All three authors contributed to the writing of the paper.   
%%  After submission of the .TEX file,                     %%
%%  you will be prompted to submit your .BBL file.         %%

% if your bibliography is in bibtex format, use those commands:
\bibliographystyle{bmc-mathphys} % Style BST file (bmc-mathphys, vancouver, spbasic).
\bibliography{RJPLJM2018}      % Bibliography file (usually '*.bib' )
% for author-year bibliography (bmc-mathphys or spbasic)
% a) write to bib file (bmc-mathphys only)
% @settings{label, options="nameyear"}

% or include bibliography directly:
% \begin{thebibliography}
% \bibitem{b1}
% \end{thebibliography}

%%%%%%%%%%%%%%%%%%%%%%%%%%%%%%%%%%%
%%                               %%
%% Figures                       %%
%%                               %%
%% NB: this is for captions and  %%
%% Titles. All graphics must be  %%
%% submitted separately and NOT  %%
%% included in the Tex document  %%
%%                               %%
%%%%%%%%%%%%%%%%%%%%%%%%%%%%%%%%%%%
\vfil\eject
%%
%% Do not use \listoffigures as most will included as separate files

\end{backmatter}
\end{document}